\documentclass[12pt]{article}

\usepackage{amsmath,amsfonts,amssymb}
\newcommand{\fixme}[1]{\textbf{FIXME: }$\langle$\textit{#1}$\rangle$}

\newcommand{\be}{\begin{equation}}
\newcommand{\ee}{\end{equation}}
\newcommand{\bea}{\begin{eqnarray}}
\newcommand{\eea}{\end{eqnarray}}
\newcommand{\ba}{\begin{array}}
\newcommand{\ea}{\end{array}}
\newcommand{\bse}{\begin{subequations}}
\newcommand{\ese}{\end{subequations}}

\def\BHe{Bekenstein-Hawking entropy}
\makeatletter \@addtoreset{equation}{section}

\makeatother
\begin{document}

\baselineskip 18pt%

\begin{titlepage}
\vspace*{1mm}%
\hfill%
\vbox{
    \halign{#\hfil \cr
\;\;\;\;\;\;\;\;\;\; IPM/P-2008/005 \cr
           \;\;\;\;\;\;\;\;\;\;\;\;\;\; SUT-P-08-1a   \cr
   \;\;\;\;\;\;\;\;\;\;\;\;\;\; IC/2008/003 \cr
%arXiv:08mm.nnnn {\tt [hep-th]} \cr
          % \cr
           } % end of \halign
      }  % end of \vbox
\vspace*{15mm}%

\centerline{{\LARGE {\bf  \textsl
{Nearing  Extremal Intersecting Giants}}}}%
\vspace{1mm}
\centerline{{\Large {\bf and}}}%\vspace*{3mm}
\vspace{1mm}
\centerline{{\LARGE {\bf \textsl
{New Decoupled Sectors in ${\cal N}=4$ SYM  }}}}%
\vspace*{15mm}
%\begin{center}
\centerline{{\bf \large{R.~Fareghbal$^{1,2}$, C.~N.~Gowdigere$^{3}$,
A.~E.~Mosaffa$^{1}$, M.~M.~Sheikh-Jabbari$^{1}$}}}
%\end{center}
\begin{center}
%\vspace*{0.4cm}
{\it {$^1$Institute for Studies in Theoretical Physics and Mathematics (IPM)\\
P.O.Box 19395-5531, Tehran, IRAN\\
$^2$Department of Physics, Sharif University of Technology\\
P.O.Box 11365-9161, Tehran, IRAN\\
$^3$ The Abdus Salam ICTP, Strada Costiera 11, 34014 Trieste, ITALY
}}

{E-mails: {\tt fareghbal, mosaffa, jabbari @theory.ipm.ac.ir, cgowdige@ictp.it}}%
%\vspace*{1.5cm}
\end{center}

\begin{center}{\bf Abstract}\end{center}
\begin{quote}
We study near-horizon limits of near-extremal charged black hole
solutions to five-dimensional $U(1)^3$ gauged supergravity
carrying two charges, extending  the recent work of
Balasubramanian et.al. \cite{Balasubramanian:2007bs}. We show that there are two near-horizon decoupling limits for the near-extremal black holes, one corresponding to the near-BPS case and the other for the far from BPS case. Both of these limits are only defined on the $10d$ IIB
uplift of the $5d$ black holes, resulting in a decoupled geometry
with a six-dimensional part (conformal to) a rotating BTZ$\times
S^3$. We study various aspects of these decoupling limits both
from the gravity side and the dual field theory side. For the
latter we argue that there should be two different, but
equivalent, dual gauge theory descriptions, one in terms of the
$2d$ CFT's dual to the rotating BTZ and the other as certain large
$R$-charge sectors of $d=4,$ ${\cal N}=4$ $U(N)$ SYM theory. We
discuss new BMN-type sectors of the  ${\cal N}=4$  SYM in the
$N\to\infty$ limit in which the engineering dimensions scale as
$N^{3/2}$ (for the near-BPS case) and as $N^2$ (for the far from BPS case).

\end{quote}
\end{titlepage}
%
%*******************************************************************************************************************************************
%
\tableofcontents
\newpage

\section{Introduction and Summary}

According to AdS/CFT conjecture \cite{AdS-CFT, MAGOO} any
state/physical process in the asymptotically $AdS_5\times S^5$
geometry can be described by a (perturbative) deformation of ${\cal
N}=4,\ d=4$ supersymmetric Yang-Mills (SYM) theory. A class of
deformations of $AdS_5\times S^5$  are solutions to ${\cal N}=2,\
d=5$ $U(1)^3$ gauged supergravity (the ``gauged STU model''), for a
review e.g. see \cite{Cvetic:1999xp,Duff:1999rk}). Among these
solutions there are geometries carrying charges under some (or all)
of the three $U(1)$'s. These are generically $5d$ black hole type
solutions. It is possible to uplift these solutions to $10d$ and
obtain the corresponding type IIB solutions which are constant
dilaton solutions only involving metric and the (self-dual)
five-form field of IIB theory. These solutions which have been
extensively studied from the gravity viewpoint (e.g. see
\cite{Duff:1999rk} and references therein) can be $1/2,\ 1/4,\ 1/8$
BPS respectively preserving $16,\ 8,\ 4$ supercharges. The $10d$ BPS
solutions have been called superstars \cite{Myers:2001aq}.

In the $10d$ picture the $1/2$ BPS solutions correspond to smeared
(delocalized) spherical D3-branes \cite{Myers:2001aq}, the giant
gravitons \cite{giant}. These are branes wrapping a three sphere
inside the $S^5$ part of the background $AdS_5\times S^5$ geometry
while moving on a  geodesic along an $S^1\in S^5$ transverse to the
worldvolume $S^3$ and smeared (delocalized) over the remaining
direction. The $1/2$ BPS solutions are specified by a single
parameter, the value of the charge. In a similar manner the
two-charge $1/4$ BPS and three-charge $1/8$ BPS solutions can be
understood as geometries corresponding to intersecting giant
gravitons. The non-supersymmetric cases then correspond to turning
on specific open string excitations on the supersymmetric
(intersecting) giant gravitons.

Besides the (excited intersecting) spherical brane picture the
$5d$ charged black hole type solutions should also have a
description in the ${\cal N}=4$ SYM on $R\times S^3$. The $1/2$
BPS case is described by chiral primary operators in the
subdeterminant basis \cite{sub-det}. In a similar fashion less BPS
solutions correspond to operators involving two or three complex
scalars in the ${\cal N}=4$ vector multiplet \cite{Ramgoolam}. The
non-supersymmetric configurations when the solution is near-BPS
(\emph{i.e.} when $\frac{\Delta-J}{J}\ll 1$, where $\Delta$ is the
scaling dimension and $J$ is the $R$-charge of the corresponding
operators) then correspond to insertion of ``impurities'' in the
subdeterminant operators \cite{vijay-etal, open-string, deMello}.

In this paper we intend to extend and elaborate further on the
discussions of \cite{Balasubramanian:2007bs} and focus on the
two-charge solutions. Noting that for these solutions we have a
simple interpretation in terms of intersecting giants we pose the
following question: Is there a limit in which the (low energy
effective) gauge theory residing on the intersecting spherical brane
system decouples from the bulk? In this paper we argue, by gathering
supportive evidence from various sides, that the answer to this
question is positive. As the first and very suggestive piece of
evidence we show that there exist \emph{two} such near-horizon,
near-extremal limits, one corresponding to the \emph{near-BPS} case and the other to the \emph{far from BPS} case. In both cases there is an $X_{M,J}\times
S^3$ geometry where $X_{M,J}$ is a global $AdS_3$ or an $AdS_3$ with
conical singularity or a (rotating) BTZ black hole.

We use the existence of these decoupled geometries and the
appearance of $AdS_3$ factors to argue that string theory on both of
these decoupled backgrounds should have a description in terms of a
$2d$ CFT, which is living on the intersection of two sets of
spherical D3-branes, intersecting on an $S^1$ (cross time).
Recalling that the geometry we start with is an asymptotically
$AdS_5\times S^5$ space-time we expect to also have a description in
terms of ${\cal N}=4$ SYM on $R\times S^3$. Explicitly, there should
be a sector (sectors) of ${\cal N}=4$ SYM which is effectively
described by a $2d$ gauge theory. We identify both the $2d$ gauge
theory and the corresponding sector in ${\cal N}=4$ SYM for  the
near-BPS decoupling limit. For the non-BPS decoupling limit we
identify the corresponding sector in ${\cal N}=4$ SYM and discuss
properties of our conjectured $2d$ dual CFT.

This paper is organized as follows. In section \ref{decoupling-limits-section}, after reviewing the $5d$
SUGRA charged black hole solutions, we identify two extremal cases,
the BPS solution and the non-BPS black hole solution which has a
null singularity. We then focus on the two-charge case, turn on a
``small'' non-extremality parameter and take the ``near-horizon''
limit about these two extremal solutions to obtain two decoupled
geometries containing $AdS_3\times S^3$ factors. In the BPS case our
solution is either supersymmetric or is a deformation of a
supersymmetric background and the deformation parameter can be
continuously tuned to zero. This case was discussed to some extent
in \cite{Balasubramanian:2007bs} but for completeness we have
included it. In the non-BPS case, the solution
obtained after taking the limit is far from being BPS.

In section \ref{BH-entropy-section}, we compute the
(Bekenstein-Hawking) entropy of the corresponding $5d$ black hole
and compare it against the same entropy for the $3d$ black hole and
find an exact matching for both the BPS and the non-BPS cases. We
take this matching as an evidence for the fact that in both cases we
have a decoupled theory. This is of particular significance
especially for the non-BPS case. In this section we make both of the
near-horizon, near-extremal limits of previous section more precise
by imposing the conditions under which one can trust the classical
gravity description of the geometry obtained after the limit.

In section \ref{6d-analysis-section}, we discuss a novel
\emph{consistent} reduction of $10d$ IIB SUGRA to a
six-dimensional (super)gravity theory which besides the metric,
involves a two-form and a scalar field with a non-trivial
potential. Moreover, we also examine the Sen's entropy function
method \cite{Sen-review} for computing the black hole entropy in
$10d$, $5d$ as well as $6d$ and $3d$ viewpoints. We show that the
$10d$ giant gravitons appear as strings, source of the two-form
field, in this reduced $6d$ theory.

In section \ref{3rd-charge-section},  we show that one can turn on
the third charge in a perturbative manner (keeping the third charge
much smaller than the other two). In this way, repeating the
near-horizon, near-extremal limits on these three-charge geometries
we obtain a rotating BTZ black hole. We again have two options,
taking the near-horizon limit on the near-BPS solution or on the non-BPS, solution. We study the associated Bekenstein-Hawking entropy of these solutions from $5d$ and $3d$
viewpoints and show that, similarly to the static case, we obtain
exactly the same result for the entropies.

In section \ref{dual-gauge-section}, we elaborate on the $2d$ and
$4d$ dual gauge theory descriptions of the decoupled near-horizon
geometries for both of the near-BPS and far from BPS cases. The
$2d$ dual gauge theory for the near-BPS case is closely related to
standard the D1-D5 systems upon two T-dualities, as in this case
the radius of the spherical giant three-branes are scaled to
infinity and  hence we are essentially dealing with two stacks of
intersecting D3-branes with worldvolume $R\times S^1\times T^2$
\cite{Balasubramanian:2007bs}. In the $4d$ language taking the
near-horizon near-BPS limit corresponds to $N\to \infty$,
$g^2_{YM}=fixed$ limit and working with the sector of operators
carrying two $R$-charges, with both of the $R$-charges and the
scaling dimension $\Delta$ of order $N^{3/2}$, while
$\Delta-\sum_i J_i\sim N$. This is a generalization of the BMN
limit \cite{BMN} to the two-charge case. The far from BPS case, however corresponds to a different sector of the ${\cal N}=4$ SYM; to the sector which is far from being BPS and in
which the scaling dimension and the $R$-charges are of order $N^2$
while taking $N\to \infty$ and a certain combination of $\Delta$
and $J^2$ scales as $N$. For the near-extremal decoupled geometry,
we argue that there should be a $2d$ dual CFT description and
identify the central charge and discuss some other properties of
this conjectured $2d$ CFT.

 In the last section we give a summary of our results, outlook
and discuss interesting open questions. In two Appendices we have
gathered some useful computations and conventions. In Appendix
\ref{Reduction-Appendix}, we show the computations proving the
consistency of the reduction of the $10d$ IIB theory to the $6d$
theory discussed in section \ref{6d-analysis-section}. In Appendix
\ref{BTZ-Appendix}, we give a concise review and fix conventions we
use for the rotating BTZ and conical $AdS_3$ spaces.

\section{Decoupling Limits of Near-Extremal $5d$
Black Holes}\label{decoupling-limits-section} In this section
after reviewing the charged black hole solutions to
five-dimensional $U(1)^3$ gauged supergravity, and their uplift to
$10d$ IIB theory,  we present two different near-horizon
decoupling limits over the near-extremal black holes carrying two
charges, one for the near-BPS solution and the other for
far from BPS configuration.

\subsection{Charged black hole solutions in $5d$}

The black hole solutions that we consider in this paper were first
obtained in the five-dimensional context in \cite{Behrndt:1998ns,
Behrndt:1998jd}. They are static charged solutions to ${\cal N} =
2$ $U(1)^3$ gauged supergravity in five dimensions and hence are
black hole solutions in the $AdS_5$ background. These solutions
can be uplifted to ten dimensions as black hole (black-brane)
deformations to $AdS_5\times S^5$  \cite{Cvetic:1999xp} (see
\cite{Duff:1999rk} for a review). We will first review the
ten-dimensional black-brane solution.\footnote{We will follow the
equations of \cite{Buchel:2006dg}, which corrects a typo in
\cite{Cvetic:1999xp}.} The metric takes the form,
\begin{equation}\label{10-dim-general}
ds_{10}^2= {\sqrt{\Delta}}\ ds^2_5+\frac{1}{\sqrt{\Delta}}\
d\Sigma_5^2
\end{equation}
where%
\bse\label{5d+5d-metrics}
\begin{align}
ds^2_5& =-\frac{f}{H_1H_2H_3}dt^2+\frac{dr^2}{f} +r^2\,d
\Omega_3^2\\
d\Sigma_5^2 &=\sum_{i=1}^3L^2
H_i\left(d\mu_i^2+\mu_i^2\left[d\phi_i+a_i\ dt\right]^2\right).
\end{align}
\ese%
$(H_1H_2H_3)^{1/3}\ ds^2_5$ is the line element for the
corresponding charged $5d$ black hole and $d\Sigma_5^2$ is the
metric for a deformed $S^5$. In the above $d \Omega_3^2$ is the
round-metric on the unit $S^3$ and %
\bse\label{Hif-10d}
\begin{align}
 H_i = 1 + \frac{q_i}{r^2}, &\qquad a_i = \frac{{\tilde q}_i}{q_i} \frac{1}{L}
 \left(
\frac{1}{H_i}-1 \right), \\
f = 1 - \frac{\mu}{r^2} + \frac{r^2}{L^2} H_1 \, H_2 \, H_3, &
\qquad \Delta = H_1 \, H_2 \, H_3
\left[\frac{\mu_1^2}{H_1}+\frac{\mu_2^2}{H_2}+\frac{\mu_3^2}{H_3}\right],\\
\mu_1 = \cos \theta_1, \qquad \mu_2 &= \sin \theta_1 \cos \theta_2,
\qquad \mu_3 = \sin \theta_1 \sin \theta_2.
\end{align}\ese%
As can be readily seen the ten-dimensional solutions asymptote
(i.e. as $r\rightarrow \infty$) to $AdS_5 \times S^5$ where  the
radii of both of the $AdS_5$ and the $S^5$ are $L$. The $S^5$ is
parameterized with the angles $\theta_1, \theta_2, \phi_1, \phi_2,
\phi_3$. In terms of the $5d$ $U(1)^3$ gauged SUGRA the three
gauge fields are given by the $a_i$ (\ref{Hif-10d}a)
\cite{Cvetic:1999xp}.

The above metric represents a solution to $10d$ type IIB SUGRA with
constant dialton and with the following RR four-form gauge field
\begin{equation}\label{five-form}
B_4 = - \frac{r^4}{L} \Delta\, dt \wedge d^3\Omega - L \sum_{i=1}^3
{\tilde q}_i \,\mu_i^2 \,\left( L\, d\phi_i - \frac{q_i}{{\tilde
q}_i} dt \right) \wedge d^3\Omega,
\end{equation}
where $d^3\Omega$ is the volume form on the unit three-sphere. The
physical five-form field strength is obtained as
\[
{\cal F}_5=F_5+\ {\ast} F_5,\qquad F_5=d B_4\ .
\]

As $5d$ black holes the above solutions are identified with the
physical ADM mass $M$ and charges $\tilde{q}_i$ which in terms of
parameters of the solution $\mu$ and $q_i$ are given by
\cite{ADM-mass-charge} \footnote{We would like to thank Alex
Buchel, Mirjam Cvetic and Wafic Sabra for useful correspondence
on the notion of (ADM) mass in the AdS backgrounds for gauged STU supergravity models.}$^,$%
\footnote{See also \cite{de-Haro} for a general discussion on the
relation between the ADM mass and charge in the holographic
setting.}
\bse\label{5d-ADM-mass-charge}%
\begin{align}
{\tilde q}_i &= \sqrt{q_i ( \mu+q_i)}\\ M& = \frac{\pi}{4 G_N^{(5)}}
(\frac32 \mu + q_1 + q_2 + q_3 +\frac{3L^2}{8}),
\end{align}
\ese%
where $G_N^{(5)}$ is the five-dimensional Newton constant and is
related to the ten-dimensional one as%
\be\label{5d-G-Newton}%
G_N^{(5)}=G^{(10)}_N\ \frac{1}{\pi^3 L^5}.
\ee%
The last term in the ADM mass expression
(\ref{5d-ADM-mass-charge}b) is the Casimir energy coming due to
the fact that the global $AdS_5$ background has a compact $R\times
S^3$ boundary. $\mu$ is a parameter which measures deviation from
being BPS. For $\mu=0$ case, $\tilde q_i=q_i$ and hence ADM mass
up to the Casimir energy and factor of $\pi/4G_N^{(5)}$ is equal
to the sum of the physical charges and therefore the solution is
BPS. The BPS configuration with $n$ number of non-vanishing
$q_i$'s ($n=1,2,3 $) generically preserves $1/2^n$ of the 32
supercharges of the $AdS_5\times S^5$ background, except for the
three-charge case with $q_1=q_2=q_3$ which is 1/4 BPS  and
corresponds to a $5d$ AdS-Reissner-Nordstrom black hole
\cite{AdS5-RN}. All the supersymmetric BPS solutions have naked
singularity.

Black holes with regular horizons can only occur  when $\mu \neq
0$ and hence are all non-supersymmetric. For the $\mu\neq 0$ cases
depending on  the number of non-zero charges, which can be one,
two or three, we have different singularity and horizon structures
\cite{Balasubramanian:2007bs, Myers:2001aq, Behrndt:1998jd}.

As ten-dimensional IIB solutions, these black holes correspond to
(smeared or delocalized) stack of intersecting spherical three-brane
giant gravitons wrapping different $S^3\in S^5$. The angular
momentum that each stack of giants carries is
\cite{Myers:2001aq}%
\be\label{Ji-giant-ang-mom}%
 J_i=\frac{\pi L}{4 G_N^{(5)}} \tilde q_i\ .%
\ee%
 The number of branes in each stack is then given by
\cite{Myers:2001aq}%
 \be\label{Ni-Ji}%
N_i=\frac{2J_i}{N}=\frac{\pi^4}{2 N}\cdot \frac{L^8}{G^{(10)}_N}\cdot \frac{\tilde q_i}{L^2}, %
\ee%
 which could be understood noting that
each giant, being a D3-brane and obeying the DBI action, is
carrying one unit of the RR charge in units of three-brane tension
$T_3=1/(8\pi^3 l_s^4 g_s)$.

Here we give a short review of cases with different number of
charges.
\begin{itemize}
\item{\textbf{One-charge black hole:}
At $\mu=0$ we have a null nakedly singular solution which
preserves 16 supercharges. As soon as we turn on $\mu$ the
solution develops a horizon with a space-like singularity sitting
behind the horizon.

The  ten-dimensional IIB uplift of these solutions contain
non-trivial  five-form flux and correspond to various giant
graviton configurations \cite{Myers:2001aq}. The one charge case
with $\mu=0$ corresponds to  1/2 BPS three sphere giant
configuration wrapping an $S^3$ inside the $S^5$ while moving with
the angular momentum $J\propto q$. This gravity configuration,
however, describes a giant smeared over (delocalized in) two
directions inside $S^5$ transverse to the worldvolume of the
brane. Turning on $\mu$ then corresponds to adding open string
excitations to the giant graviton while keeping the spherical
shape of the giant.}
\item{
\textbf{Two-charge black hole:} For $0\leq \mu<\mu_c$ we have a
time-like but naked singularity where $\mu_c=q_2q_3/L^2$. At
$\mu=\mu_c$ we have an \textit{extremal, but non-BPS} black hole
solution with a zero size horizon area (horizon is at $r=0$) and
$r=0$ in this case is a null naked singularity. As we increase $\mu$
from $\mu_c$ the solution develops a finite size horizon and the
space-like singularity hides behind the horizon.

As ten-dimensional solutions, the two-charge case at $\mu=0$
corresponds to two sets of delocalized giant gravitons wrapping
two $S^3$'s inside $S^5$ while rotating on two different $S^1$
directions. The worldvolume of the giants  overlap on a circle. If
one of the charges is much smaller than the other one a better
(perturbative) description of the system is in terms of a rotating
single giant where as a result of the rotation  the giant is
deformed from the spherical shape. As in the single charge case,
turning on $\mu$, especially when $\mu$ is small enough,
corresponds to adding open string excitations while keeping the
$U(1)$ symmetry of the giants intersection.

For the extremal case at $\mu=\mu_c$ the brane picture is more
involved. In this case we are dealing with intersecting giants which
are generically far from being BPS and effectively we are dealing
with a stack of giants with worldvolume $R\times S^1\times
\Sigma_2$, where $\Sigma_2$ is a compact $2d$ surface inside the
$S^5$. Out of extremality, measured by $\mu-\mu_c$, then corresponds
to excitations/fluctuations above this stack of giants. In the rest
of this paper we will study dynamics of a class of these
excitations.}

\item{
\textbf{Three-charge black hole:} For $0\leq \mu<\mu_c$ we have a
time-like naked singularity, the singularity is, however, behind
$r=0$ (one can extend the geometry past $r=0$). At some critical
$\mu$, $\mu=\mu_c$, we have an extremal solution with a finite size
horizon (function $f$ has double zeros at some $r_h\neq 0$)
\cite{Behrndt:1998jd}. For $\mu>\mu_c$ the geometry has two inner
and outer horizons.

From the ten-dimensional viewpoint the three-charge case corresponds
to a set of three smeared giant gravitons intersecting only on the
time direction and the giants in each set moving on either of the
three $S^1$ directions in the $S^5$, which in (\ref{5d+5d-metrics}b)
are parameterized by $\phi_i$. Again if one of the charges is much
smaller than the other two a better description of the system is in
terms of two giants intersecting on an $S^1$, but the third charge
appears as a rotation on the $S^1$. We will return to this latter
case in more detail in section \ref{3rd-charge-section}.}
\end{itemize}

\subsection{The near-horizon limit of two-charge  solutions}

As discussed and reviewed in the previous subsection for the two and
three-charge cases we have extremal black holes. These extremal
black holes can be BPS or  non-BPS. One may then expect that for
both of these cases there should exist a ``near-horizon'' limit in
which the theory on the corresponding intersecting giants decouple
from the bulk. To study this we need to first analyze the
near-horizon geometry for such extremal (or near-extremal)
solutions. Although in this paper we mainly focus on the two-charge
case, we  discuss the three-charge case, when the third charge is
much smaller than the other two, in section
\ref{3rd-charge-section}. We analyze and discuss both of the
two-charge near-BPS and far from BPS cases in parallel.
The near-BPS case has also been analyzed in
\cite{Balasubramanian:2007bs}.

To start the analysis let us choose the two non-vanishing charges to
be $q_2$ and $q_3$. In
this case the function $f$ in the metric takes the form%
\be\label{two-charge-f}%
f=\frac{r^2}{L^2}+f_0-\frac{\mu-\mu_c}{r^2},
\ee%
where
\be\label{f0}%
 f_0=1+ \frac{q_2+q_3}{L^2},%
\ee%
and
\be\label{muc}%
 \mu_c=\frac{q_2q_3}{L^2}.%
\ee%

We use the $5d$ metric to locate the horizon, which occurs where
$g^{rr}$ vanishes, or at the roots of $r^{4/3}f$. From
\eqref{two-charge-f} it is evident that for $\mu=\mu_c$ we have a
double zero at $r=0$ (for $\mu<\mu_c$ $f$ is positive definite and
for $\mu>\mu_c$ $f$ has a single positive root). Therefore, at
$\mu=\mu_c$ we are dealing with an extremal black hole (or from
$10d$ viewpoint, black-brane) solution in which both horizon and
singularity are at $r=0$.

 The radius of the $S^3$ in the five-dimensional metric is proportional to $( H_1
H_2 H_3)^{\frac13} r^2$. Only for the three-charge case, in the
near-horizon limit $r \rightarrow r_h \neq 0$, we get a constant
term \cite{Morales:2006gm}. For the two-charge case, the
near-horizon limit $r \rightarrow 0$ gives $(q_2 q_3)^{\frac13}
r^{\frac23}$ which is clearly not a product geometry. As we will
show, however, the factorization happens if we take the limit over
the $10d$ solution and this is what we do here.

\subsubsection{Near-horizon limit, the near-BPS case}\label{NH-BPS}
As argued the BPS case happens when $\mu=0$. In the near-horizon
limit we consider in this subsection, together with $r\to 0$ we also
consider $\mu$ to be very small, explicitly we consider
either of the following limits \cite{Balasubramanian:2007bs}\\%

 $\bullet$ \emph{$\mu_1\sim 1$ case}
\begin{equation}\label{the-limit-BPS-1}
\begin{split}
r = \epsilon \tilde\rho, &\qquad \mu_i=
\epsilon^{1/2} x_i,\\
\mu-\mu_c= \epsilon^2 M , &\qquad q_i=\epsilon \hat q_i , \  i=2,3,
\end{split}
\end{equation}
while keeping $\tilde\rho,\ \hat q_i,\ M,\ x_i, \phi_i, \ L$
fixed. Note also that, as $\mu_1^2=1-\mu_2^2-\mu_3^2$, in this
limit $\mu_1=1+{\cal O}(\epsilon^2)$. This limit corresponds to
$\theta_1\sim \epsilon^{1/2}, \theta_2=$fixed \emph{cf.}
(\ref{Hif-10d}c).\\

$\bullet$ \emph{$\mu_1\sim \mu_1^0\neq 1$ case}
\begin{equation}\label{the-limit-BPS-2}
\begin{split}
r = \epsilon \tilde\rho, &\qquad \theta_i=\theta_i^0-\epsilon^{1/2}
\hat \theta_i, \ 0\leq\theta_i^0\leq \pi/2, \\
   \mu-\mu_c= \epsilon^2 M , &\qquad q_i=\epsilon \hat
q_i, \qquad
\psi_i=\frac{1}{\epsilon^{1/2}}\left(\phi_i-\frac{t}{L}\right),\
i=2,3,
\end{split}
\end{equation}
while keeping $\tilde\rho,\ \hat q_i,\ M,\ \theta_i^0,\ x_i, \ L$
fixed.

As we can see in both of these cases%
\be\label{deficit}%
\gamma^2\equiv \frac{\mu-\mu_c}{\mu_c}%
\ee%
is kept fixed,  $\mu\sim \epsilon^2$ and hence the physical charges
$\tilde q_i=q_i\sim \epsilon$.

Taking the limit we arrive at the $AdS_3\times S^3\times T^4$
geometry%
\be\label{metric-BPS-limit}%
ds^2=\epsilon\left[
R^2_S\left(ds^2_{AdS}+d\Omega_3^2\right)+\frac{L^2}{R^2_S}
ds^2_{{\cal C}_4} \right]\ee %
where
\be%
ds^2_{AdS}=-(\rho^2-\gamma^2)
d\tau^2+\frac{d\rho^2}{\rho^2-\gamma^2}+\rho^2d\phi_1^2%
\ee%
with
\[ \rho=\frac{L}{(\hat q_2\hat q_3)^{1/2}} \frac{r}{\epsilon},\qquad \tau=\frac{1}{L }t.\]%

The $S^3$ radius $R^2_S$ and the four-dimensional metric
$ds^2_{{\cal C}_4}$ have different forms for the two cases:

 $\bullet$ \emph{$\mu_1\sim 1$ case}
\be\label{theta1=0} %\begin{align}%
R^2_S=\sqrt{\hat q_2\hat q_3}, \qquad ds^2_{{\cal C}_4}=\sum_{i=2,3}
\hat q_i (dx_i^2+x_i^2d\psi_i^2)%
%\end{align}
\ee%
where $\psi_i=\phi_i-\frac{t}{L}$.

$\bullet$ \emph{$\mu_1\sim \mu_1^0\neq 1$ case}
\be\label{fixed-theta} %\begin{align}%
R^2_S=\sqrt{\hat q_2\hat q_3}\mu_1^0, \qquad ds^2_{{\cal
C}_4}=\sum_{i=2,3} \hat q_i (dx_i^2+(\mu_i^0)^2d\psi_i^2)%
%\end{align}
\ee%
where $\mu_2^0=\sin\theta_1^0\cos\theta_2^0,\
\mu_3^0=\sin\theta_1^0\sin\theta_2^0$,
$dx_2=\cos\theta_1^0\cos\theta_2^0d\hat\theta_1$ and
$dx_3=\cos\theta_1^0\sin\theta_2^0d\hat\theta_1+\cos\theta_2^0\sin\theta_1^0d\hat\theta_2$.

In either case the ${\cal C}_4$ part of the geometry after
appropriate periodic identifications is describing a $T^4$ and hence
the solutions are $AdS_3\times S^3\times T^4$. For $\gamma^2=-1$ we
have a global $AdS_3$ space, for $-1<\gamma^2<0$ it is a conical
space, for $\gamma^2=0$ we have a massless BTZ and for $\gamma^2>0$
we are dealing with a static BTZ black hole of mass $\gamma^2$. (For
more detailed discussion see Appendix \ref{BTZ-Appendix}.) These
geometries are, upon two T-dualities, related to standard the D1-D5
system and the corresponding arguments are applicable to this case
\cite{Balasubramanian:2007bs, Maldacena:2000dr}. A detailed
discussion on the $AdS_3\times S^3$ geometries and the spectrum of
supergravity/string theory in $AdS_3\times S^3$ compactification may
be found in \cite{de-Boer-6d} and references therein. We will give a
brief review in section \ref{2d-CFT-NBPS-section}.

\subsubsection{Near-horizon limit, the far from BPS case}\label{non-BPS-extremal-limit}%
 We take the following near-horizon decoupling limit over the \textit{far from BPS} solution,
while \emph{keeping $\mu_c$ fixed}, \emph{i.e.}
\begin{equation}\label{the-limit-static}
\begin{split}
r = \epsilon \tilde\rho, &\qquad  t =
\frac{\tilde\tau}{\epsilon}, \qquad \mu-\mu_c= \epsilon^2 M\\
\qquad \phi_1 = \frac{\varphi}{\epsilon}, &\qquad \phi_i = \psi_i
+ \frac{{\tilde q}_i}{q_i L} \frac{\tilde\tau}{\epsilon},\ i=2,3
\end{split}
\end{equation}
where $\tilde\rho, \ \tilde\tau,\ \varphi,\ \psi_i,\ M, \ L$ are
kept fixed while taking $\epsilon\to 0$. Taking this limit we also
keep $q_i/L^2$ and hence $f_0$, $\mu_c/L^2$ fixed.

In the above near-horizon near-extremal limit, the leading
contribution from functions $f$, $\Delta,\ H_i$ appearing in
\eqref{10-dim-general} become
\begin{equation}\label{f-Delta-limit}
f = f_0 - \frac{M}{\tilde\rho^2}, \qquad \Delta=\mu_1^2\
\frac{q_2q_3}{\tilde \rho^4}\cdot\frac{1}{\epsilon^4},\qquad
H_i=\frac{q_i}{\tilde\rho^2}\cdot\frac{1}{\epsilon^2}\ .
\end{equation}
The ten-dimensional metric \eqref{10-dim-general} in the limit
\eqref{the-limit-static}, after some redefinition of coordinates
takes the form
\begin{equation}\label{decoupled-metric-static}
ds^2_{10} = \mu_1 ~( R^2_{AdS_3}\ ds^2_3 +  R^2_S ~ d\Omega_3^2\,)
+ \frac1\mu_1 ds^2_{{\cal M}_4}
\end{equation}
where
\begin{equation}\label{BTZ-static}
ds^2_3 = - (\rho^2 -\rho_0^2)d\tau^2 +\frac{d\rho^2}{\rho^2
-\rho_0^2}+\rho^2d\varphi^2,
\end{equation}
$d\Omega_3^2$ is the metric for a round three sphere of unit radius
and
\begin{equation}\label{M4-metric}
ds^2_{{\cal M}_4} = \frac{L^2}{R^2_S} \left[ q_2 ~( d\mu_2^2 +
\mu_2^2\, d\psi_2^2 ) \,+ \,q_3 ~( d\mu_3^2 + \mu_3^2 \,d\psi_3^2
) \right].
\end{equation}
In the above%
\bse\label{Rs-RAdS}
\begin{align}
R^2_S\equiv\sqrt{q_2q_3}=\sqrt{L^2\mu_c},&\qquad
R^2_{AdS_3}=\frac{R^2_S}{f_0},\\
\rho_0^2=\frac{M}{\mu_c}\ ,
\end{align}\ese
and the new coordinates $\rho$ and $\tau$ in terms of the original
coordinates $t, r$ are defined as \footnote{This scaling is a
generic feature of near-horizon, near-extremal limits, \emph{e.g.}
see \cite{Bardeen:1999px}.}%
\be\label{rho-tau-t-r}%
\tau={\epsilon}\ \frac{R_S}{R_{AdS_3}}\frac{t}{L},\qquad
\rho=\frac{L}{R_SR_{AdS_3}} \frac{r}{\epsilon}\ .%
\ee%
Note that $\mu_1^2=1-\mu_2^2-\mu_3^2$ and therefore $\mu_1$ is not a
constant (in contrast to the near-BPS case). As we see after the
decoupling limit the metric has taken the form of a six-dimensional
part which is conformal to $AdS_3\times S^3$ and a four-dimensional
part conformal to ${\cal M}_4$ which is a K\"ahler manifold.

For $\rho_0^2\geq 0$ the  metric \eqref{BTZ-static} describes a
stationary BTZ black hole in a \emph{locally} \footnote{Note that
the angle $\varphi$ is ranging over $[0,2\pi\epsilon]$.} $AdS_3$
background of  radius $R_{AdS_3}$ (\ref{Rs-RAdS}a) and of mass
$\rho^2_0$ (\ref{Rs-RAdS}b).
 For $\rho^2_0<0$, however, we have
an $AdS_3$ with conical singularity and the deficit angle
$2\pi(1-\delta)$ where
%(\emph{i.e.}, $\varphi$ is ranging over $[0,2\pi\delta]$)%
\be\label{deficit-2}%
\delta=\frac{\mu-\mu_c}{\mu_c}=\epsilon^2 \rho^2_0.%
\ee%

It is notable that the angle in the BTZ which is parameterized by
$\varphi$ is coming from the part which was in the $S^5$ part of the
original $AdS_5\times S^5$ background, while the rest of the
six-dimensional part of metric come from the original $AdS_5$
geometry; the ${\cal M}_4$ is coming from the $S^5$ piece. As
mentioned the angle $\varphi$ is ranging over $[0,2\pi\epsilon]$,
nonetheless the causal boundary of the near-horizon decoupled
geometry is still $R\times S^1$. To see this we note that at large,
but fixed $\rho$
the $AdS_3$ part of the metric takes the form%
\be\label{boundary-near-Ext}%
ds_3^2\sim R^2_{AdS_3}\epsilon^2 \rho^2 (-dt^2+d\phi_1^2)\ , %
\ee%
where $t$ is the (global) time direction in the original $AdS_5$
geometry, and therefore the causal boundary of this space is
$R\times S^1$.

Metric \eqref{decoupled-metric-static} is a constant dilaton
solution to IIB SUGRA with the  four-form field
\begin{equation}\label{four-form-limit}
B_{4} = -L^2\,\left( {\tilde q}_2\, \mu_2^2~ d\psi_2 +\,{\tilde
q}_3 \ \mu_3^2~ d\psi_3\right) \wedge d^3\Omega_3,
\end{equation}
where in the near-horizon, far from BPS limit
\eqref{the-limit-static}%
\be\label{tilde-q-Next-limit}
 \tilde q_2^2=q_2^2(1+\frac{q_3}{L^2}),\qquad \tilde
q_3^2=q_3^2(1+\frac{q_2}{L^2}). %
\ee%
 The above four-form can be
obtained from taking the decoupling limit \eqref{the-limit-static}
over the four-form of the original solution \eqref{five-form}.

It is interesting to note the similarities between this decoupled
geometry and the one given in
\eqref{metric-BPS-limit}$-$\eqref{fixed-theta}. The geometry
\eqref{decoupled-metric-static} is the much expected ``global
decoupled solution'' of \cite{Balasubramanian:2007bs}. In
\eqref{metric-BPS-limit} the radii of the $AdS_3$ and the $S^3$ are
equal while in \eqref{decoupled-metric-static} they are different.
Moreover, the range of the $\varphi$ coordinate in
\eqref{decoupled-metric-static} is $[0,2\pi\epsilon]$ while that of
$\phi_1$ in \eqref{metric-BPS-limit} is $[0,2\pi]$. As a side
comment we note that, similarly to the original two-charge extremal
solution, the geometry obtained in the near-horizon far from BPS
limit, even when $\mu=\mu_c$, is not preserving any supersymmetries
of the $10d$ IIB theory.

\section{Entropy of The Two-Charge Solution}\label{BH-entropy-section}

In this section, we first compute the entropy of the $5d$ black
hole and take both the near-horizon limits on it. Moreover, we
argue how $\epsilon$ should scale with $N$ in both cases. We then
compute the entropy of the $3d$ BTZ black hole that is part of the
geometry of the near-extremal near-horizon limit, and show that it
precisely agrees with the $5d$ entropy in the same limit. This
provides the first piece of evidence for the fact that this limit
is indeed a nice decoupling limit.

\subsection{Black hole entropy, $5d$ viewpoint}\label{BH-entropy-5d}

To compute the Bekenstein-Hawking entropy of the two-charge $5d$
black hole we recall that its metric is given by
\begin{equation}\label{5d-metric}
ds^2 = - (H_2\,H_3)^{-\frac23}\ f dt^2 +
(H_2\,H_3)^{\frac13}({f}^{-1}dr^2+r^2 d^2\Omega_3).
\end{equation}
Zero(s) of $r^{4/3}f$ determine the location of the horizon(s). The area of the horizon is then
\be\label{A5-generic}
A_h^{(5)}=2\pi^2 r^3_h (H_2H_3)^{1/2}|_{r=r_h}\ .
\ee
The Bekenstein-Hawking entropy is
\[
S_{BH}=\frac{A_h^{(5)}}{4 G^{(5)}_N}\ .
\]
Recalling that
$
G^{(5)}_N=\frac{G^{(10)}_N}{\pi^3 L^5}\ ,
$
and that
 \be\label{ls-lp-L}
G^{(10)}_N=8\pi^6 g_s^2 l_s^8, \qquad  L^4=4\pi g_s N l_s^4\ ,%
\ee%
we obtain%
\be\label{BH-entropy-5d-general}%
S_{BH}=\frac{1}{2\pi} N^2\cdot \frac{A_h^{(5)}}{L^3}.%
\ee %
As we see, once the area is measured in $AdS_5$ units $L$, the
entropy generically scales like $N^2$. However, one should
remember that the area of the horizon also scales with $\epsilon$
and in fact in two different ways for the two near-horizon limits.
Therefore, we discuss the near-BPS and far from BPS cases
separately. Before that, we should stress that the notion of black
hole entropy is only valid when horizon area is not Planckian and
we are in the regime we can trust classical gravity description,
explicitly that is when
\be\label{validity-limit}%
 S_{BH}\gg 1,\qquad \frac{G^{(10)}_N}{l_s^8}\cdot
S_{BH}\gg 1 \qquad (or\ \  g^2_s\cdot S_{BH}\gg 1).%
\ee%
Moreover, one should ensure that all the curvature invariants remain small (in Planck or string units).

\subsubsection{$5d$ black hole entropy of near-BPS case}\label{BH-entropy-BPS-section}

In the near-BPS limit the horizon is located at \be\label{rh-BPS}
r^2_h=\mu-\mu_c \ee and hence \be\label{entropy-near-BPS}
S_{BH}^{Near-BPS}=\pi \gamma\frac{\hat \mu_c}{L^2}\
N^2\epsilon^2\ , \ee where $\gamma$ is defined in \eqref{deficit},
and $\hat \mu_c=\mu_c/\epsilon^2$. In this case the curvature
components scale as $1/\epsilon$ (in units of $L^{-2}$). Validity
of classical gravity arguments then implies that one should scale
$N$ to infinity as well:
$%\be\label{Nvs.epsilon-BPS}
N\sim \epsilon^{-\alpha}, \ \alpha\geq 1.
$%\ee
\ This consideration is, however, not strong enough to fix $\alpha$.
Noting the form of metric, that it has a factor of $\epsilon$ in
front and that one expects the string scale to be the shortest
physical length leads to \be\label{Nvs.epsilon-BPS-preferred}
\epsilon\sim l_s^2 \Rightarrow \quad N\sim \epsilon^{-2}\ .
\ee%
Once the above scaling of $\epsilon$ and $N$ is  considered, we see
that the entropy \eqref{entropy-near-BPS} scales as $N\sim
\epsilon^{-2}$ to infinity.

As was argued in \cite{Balasubramanian:2007bs}, only a certain class
of massless open string modes on the intersecting giants survive the
scaling \eqref{Nvs.epsilon-BPS-preferred}. We will discuss in
section \ref{dual-gauge-section} that these modes constitute the
degrees of freedom of certain $2d$ CFT's.

In sum, our complete near-horizon, near-BPS limit is defined as an
$\alpha'=l_s^2\sim \epsilon \to 0$ limit together with
\eqref{the-limit-BPS-1} (or \eqref{the-limit-BPS-2}), while keeping
$L^4\sim N l_s^4$ fixed.

\subsubsection{$5d$ black hole entropy of the far from BPS case}\label{BH-entropy-nonBPS-section}

In the far from BPS limit of \eqref{the-limit-static} to order
$\epsilon$,  we have%
\be\label{horizon-static}%
r^2_h=\frac{\mu-\mu_c}{f_0}+{\cal O}(\epsilon^4).%
\ee%
and hence%
\be\label{BH-entropy-5d-Ext}%
S_{BH}^{Far from BPS}=\pi \frac{\mu_c}{L^2}\cdot \frac{\rho_0}{\sqrt{f_0}}\  N^2\epsilon.\ %
\ee%
 To ensure \eqref{validity-limit} and also demanding the curvature components to remain small (in $10d$ string or Planck units)
one should also send $N\to \infty$ while keeping
$\rho_0$ and $\mu_c/L^2$ finite. This is done if we scale
$
N\sim \epsilon^{-\beta},\ \beta\geq \frac12\ .
$%\ee
\ In our case, as we will discuss in section \ref{6d-analysis-section}, $\beta=1$ is giving the appropriate choice,
\be\label{Nvs.epsilon-Ext}
N\sim \epsilon^{-1}\to \infty\ .
\ee

In sum, in our limit we keep $L, g_s$, $q_i/L^2$ and $\rho_0$
finite while taking $l^4_s\sim N^{-1}\sim \epsilon  \to 0$.
Similarly to the near-BPS case of section
\ref{BH-entropy-BPS-section}, in this case $S_{BH}\sim N\to
\infty$.

%It is notable that the decoupled geometry
%\eqref{decoupled-metric-static} has a curvature singularity at
%$\mu_1=0$. In our analysis we assume that we are working away from
%$\mu_1=0$. We expect this singularity to be resolved by stringy
%corrections.

\subsection{The $3d$ BTZ black hole entropy, the far from BPS
case}\label{3d-entropy-NExt-section}

To work out the \BHe\  of the BTZ black hole obtained after taking
the limit we should have the relevant three dimensional Newton
constant. To this end, we show that there is a consistent reduction
of the $10d$ IIB theory over ${\cal M}_4$ to a six-dimensional
(super)gravity theory.  Computations showing the consistency of the
reduction are given in the
Appendix \ref{Reduction-Appendix}. The Newton constant of this six-dimensional theory is \eqref{GN-6d-App}%
\be\label{GN-6}%
G^{(6)}_N=\frac{G^{(10)}}{\frac{\pi^2}{2} L^4}\ = \frac{\pi^2 L^4}{N^2},%
\ee%
and its action is given in \eqref{6d-action}.

As will be shown in the next section, the geometry we obtain after
taking the limit is BTZ$\times S^3$ solution to this $6d$ theory.
One can hence make a further reduction of this $6d$ theory on the
$S^3$ to obtain a $3d$ gravity theory. Similarly to the standard
case which \emph{e.g.} was discussed in \cite{Maldacena:2000dr},
this $3d$ (gauged super-)gravity has $SL(2,R)^2$ gauge symmetry (for
the pure gravity) and a gauge group which has $U(1)_L\times U(1)_R$
as its sub-group. Noting that in our case the radius of $S^3$ is
$R_S$,
the corresponding $3d$ Newton constant is%
\be\label{G3N}%
G^{(3)}_N=\frac{G^{(6)}_N}{2\pi^2 R^3_S}=\frac{G^{(10)}_N}{\pi^4 L^4
R^3_S}=\left(2{N^2}\cdot \frac{R^3_S}{L^4}\right)^{-1}
\ .%
\ee
The \BHe\  of the corresponding BTZ is then given by%
\[
S^{(3)}_{BH}=\frac{A^{(3)}}{4G^{(3)}_N}
\]
where $A^{(3)}$ is the area of horizon for the BTZ black hole. For the far from BPS case that is%
\be\label{3d-horizon-area}%
A^{(3)}={2\pi\epsilon} R_{AdS_3} \rho_0. %
\ee%
In computing the area of the horizon of the BTZ black hole
\eqref{3d-horizon-area} one should recall that  $\varphi$ which
parameterizes the horizon circle is ranging from $0$ to
$2\pi\epsilon$
\eqref{the-limit-static}. Therefore,%
 \be\label{3d-entropy}%
S^{(3)}_{BH}=\pi \frac{R_{AdS} R_S^3}{L^4}\ \rho_0 N^2\epsilon\ ,
\ee%
which is exactly the same as \eqref{BH-entropy-5d-Ext} once we recall that $R_{AdS}=R_S/\sqrt{f_0}$ and that
$\mu_c=R^4_S/L^2$.

The exact matching of the entropies of the $5d$ black hole and
that of the $3d$ BTZ is a strong sign of the fact that in the
decoupling \emph{far from BPS}, near-horizon limit we have taken
we have not lost any degrees of freedom.\footnote{It is noteworthy
that the horizon of the $5d$ black hole is $S^3\in AdS_5$ while
that of the $3d$ BTZ black hole is an $S^1\in S^5$ in the original
$AdS_5\times S^5$ geometry and hence is not present in the $5d$
black hole picture.} This brings the hope that despite the lack of
supersymmetry we may still look for a dual $1+1$ dimensional gauge
theory descriptions. We will return to this point is section
\ref{dual-gauge-section}.

\section{The $6d$ Analysis of the \emph{Far from BPS} Solution}\label{6d-analysis-section}

In previous sections we discussed two near-horizon decoupling
limits of the two-charge $10d$ black-brane solutions. For the
near-BPS case it is immediate to check that the $AdS_3\times
S^3\times {\cal C}_4$ obtained after the limit is again a solution
to IIB theory. This is not, however, obvious for the far from BPS
case. In this section we discuss this issue through running (and
in fact generalizing and extending) Sen's entropy function method
\cite{Sen-review} for the $6d$ BTZ$\times S^3$ geometry. In
Appendix \ref{Reduction-Appendix}, we show that there is a
consistent reduction of the $10d$ IIB theory to a $6d$ theory
described in \eqref{6d-action}; hence showing that the BTZ$\times
S^3$ is a solution to this $6d$ theory is enough to guarantee that
the $10d$ near-horizon far from BPS geometry is a solution to IIB
theory.

In addition, using the entropy function method we compute the
entropy of the near-extremal BTZ$\times S^3$ solution as the
near-horizon limit of a (extremal) black string solution of this
$6d$ theory and show that this entropy is exactly equal to the
entropy of the $5d$ black hole computed in section
\ref{BH-entropy-nonBPS-section}. This is very suggestive that one
may use this $6d$ (black) string picture to identify the dynamical
degrees of freedom of the dual $2d$ CFT description.

To run the entropy function machinery we start with our $6d$ (super)gravity
action \eqref{6d-action}:
\begin{equation}\label{6d-action-section3.3}
    S=\frac{1}{16\pi G^{(6)}_N}\int\,dx^6\sqrt{-g^{(6)}}\left[
   R^{(6)}- g^{\mu\nu}\frac{\nabla_\mu X \, \nabla_\nu X}{X^2}+\frac{4}{L^2}\big(X+\frac1X\big)-\frac{1}{3}X^2
   F_{\mu\nu\rho}F^{\mu\nu\rho}\right].
\end{equation}
The next step is the near-horizon field configuration, which for
extremal black holes (and black strings) is an $AdS_m \times S^n$
geometry. In our case, however, we have a BTZ$\times S^3$ solution.
Recalling that the Riemann curvature for a (rotating) BTZ black hole
(which is the most general $3d$ black hole geometry) is the same as
the Riemann curvature for $AdS_3$, we may use the standard steps of
the usual entropy function method. We will comment more on this
issue in the discussion section.

The ans\"atz for the field configuration is
\begin{subequations}\label{ent-func-ansatz}
\begin{align}
    ds^2 &=v_1\Big(-(\rho^2-\rho_0^2)d\tau^2+\frac{d\rho^2}{\rho^2-\rho_0^2}+\rho^2d\varphi^2\Big)+\,v_2 d\Omega_3^2\\
    %\Big(d\theta_1^2+sin^2\theta_1(d\theta_2^2+sin^2\theta_2d\psi^2)\Big)\\
    F &=e\ vol(AdS_3)+ p\ vol(S^3)\\
      X & =u=const,
\end{align}
\end{subequations}
where $vol(AdS_3)$ and $vol(S^3)$ are respectively the volume forms
of $AdS_3$ (or BTZ black hole) and the three sphere of unit radius.
$v_1$, $v_2$, $e$ and $u$ are constants to be determined by the
equations of motion in terms of the electric charge $Q_e$,
\be\label{eletrci-charge-def}%
Q_e\equiv\frac{1}{8\pi^2} \int_{S^3}\frac{\partial (\sqrt{-g^{(6)}}{\cal L})}{\partial F_{\tau\rho\varphi}} d^3\Omega\,%
\ee%
where the above integration over an $S^3$ of unit radius,  and the
magnetic charge $Q_m$ which is equal to $p$. The equations
governing these parameters are obtained from variation of the
\emph{entropy function} $F$ which is defined as
\begin{equation}\label{F-def}
\begin{split}
    F(v_i,e,u; Q_e, p=Q_m)&=\frac{1}{16 G^{(6)}_N}\int\,dx^H\,\sqrt{-g^{(6)}}\Big(\frac12F_{\tau\mu\nu}
    \frac{\partial \mathcal{L}}{\partial F_{\tau\mu\nu}}-
    \mathcal{L}\Big),
    %\cr&=
    %\frac{1}{4 G^{(6)}_N}\int\,dx^H\,\sqrt{-g^{(6)}}\Big({v_1^{-\frac32} v_2^{-\frac32}}e Q_e- \frac14\mathcal{L}\Big),
\end{split}
\end{equation}
the $\frac12$ factor  has appeared because we are dealing with a
two-form field and $\{x^H\}$ is the four-dimensional horizon of
the $6d$ (presumably) black string solution. For us and in metric
(\ref{ent-func-ansatz}a), this is $S^1 \times S^3$ where $S^1$ is
a circle of radius $\rho_0$ parameterized by $\varphi\in [0, 2 \pi
\epsilon]$. \footnote{It should be noted that gravity equations of
motion, and hence the entropy function method, are local
differential equations and are hence  blind to the range of
coordinates \emph{e.g.} the $\varphi$ direction.}

Plugging the ans\"atz \eqref{ent-func-ansatz} into \eqref{F-def}
we find
\begin{equation}\label{F-plugged}
\begin{split}
    F(v_i,e,u; Q_e,
    Q_m)=\frac{\pi^3\rho_0\epsilon}{ G^{(6)}_N}\Bigg(eQ_e&
    -\frac14(v_1v_2)^\frac32\left[\frac{6}{v_2}-\frac{6}{v_1}
        +\frac{4}{L^2}\big(u+\frac1u\big)\right]+\\
    &+\frac12u^2\left[Q_m^2\left(\frac{v_1}{v_2}\right)^\frac32-e^2\left(\frac{v_2}{v_1}\right)^\frac32\right]\Bigg).
\end{split}\end{equation}%
 Field equations which give values of $v_i,u, e$
in terms of electric and magnetic charges $Q_e$ and $Q_m$ are
\begin{equation}\label{uve-QQ}%\begin{align}
\frac{\partial F(v_i,e,u; Q_e,Q_m)}{\partial \Phi_I} =0, \qquad\qquad \Phi_I=\{v_i, e, u\}\\
%\frac{\partial F(v_i,e,u; Q_e, Q_m)}{\partial e} &=0,\\
%  \frac{\partial F(v_i,e,u ; Q_e, Q_m)}{\partial u} &=0. \end{align}
\end{equation}
After some simplifications the above equations take the form
\begin{subequations}\label{ent-func-eqn}\begin{align}
    \frac{1}{v_2}-\frac{1}{v_1}&=-\frac{1}{L^2}\Big(u+\frac1u\Big), \\
    \frac{1}{v_2}+\frac{1}{v_1}&=u^2\Big(\frac{e^2}{v_1^3}+\frac{Q_m^2}{v_2^3}\Big), \\
    \frac{1}{L^2}\Big(u-\frac1u\Big)&=-u^2\Big(\frac{e^2}{v_1^3}-\frac{Q_m^2}{v_2^3}\Big), \\
    Q_e&=\Big(\frac{v_2}{v_1}\Big)^\frac32u^2e.
\end{align}\end{subequations}
It is readily seen that%
\be\label{ent-func-solution}%
v_2=\sqrt{q_2q_3},\qquad
v_1=\frac{\sqrt{q_2q_3}}{1+\frac{q_2+q_3}{L^2}}, \qquad
u=\sqrt{\frac{q_2}{q_3}}\ , %
\ee %
is a solution to
\eqref{ent-func-eqn} provided that %
\be\label{tildeq-QQ-1}%
Q_e^2 \equiv q_2^2(1+\frac{q_3}{L^2}), \qquad\quad Q^2_m\equiv
q_3^2(1+\frac{q_2}{L^2})\ . %
\ee%
In order to match our notations and conventions with those of
previous sections we choose: %
\be\label{tildeq-QQ}%
Q_e \equiv \tilde
q_2, \qquad\quad Q_m\equiv
\tilde q_3\ . %
\ee%
The above provides a crosscheck that the BTZ$\times S^3$ geometry
obtained is indeed a solution to our $6d$ theory and hence the $10d$
IIB theory. Moreover, it makes a direct connection between the $6d$
charges $Q_e, Q_m$ and the number of five-form fluxes (and hence
number of giants) in the $10d$ setting.

As discussed in section \ref{decoupling-limits-section}, in the
$10d$ setting in the two-charge case we are dealing with a system
of intersecting giant three-sphere branes consisting of $N\tilde
q_2/2L^2$ and $N\tilde q_3/2L^2$ giants which are intersecting
over a circle (the $\varphi$ direction in our BTZ$\times S^3$
solution). In the $6d$ setting, however,  we are dealing with a
system of strings along $\varphi$ direction which are electrically
and magnetically charged under the three-form $F_3$. These
(circular) strings are the giant three-branes wrapping two cycles
of the four-dimensional reduction manifold ${\cal M}_4$, one set
of them are wrapping $\mu_2,\psi_2$ directions and the other
$\mu_3,\psi_3$ directions.
The tension of the $6d$ strings are then %
\be\label{6d-string-tension}%
T_s^{(6)}=T_3 ({\pi L^2})=\frac{N}{2\pi L^2}\ ,%
\ee%
where we have used $T_3^{-1}=(2\pi)^3 l_s^4 g_s$, $L^4=4\pi l_s^4
g_s N$ and that the volume of the two cycles over which the
three-branes are wrapping, are $\pi L^2$ (see also
\cite{Gubser:2004xx}
for a similar discussion). It is also notable that%
\be%
T^{(6)}_s= \frac{1}{2\sqrt{G^{(6)}_N}}. %
\ee%

In the near-horizon far from BPS limit as discussed in previous
sections, we are taking $N\to \infty$ and this is done in a such a
way that (\emph{cf.} \eqref{Nvs.epsilon-Ext}) $T^{(6)}_s\sim
\epsilon^{-1}$. In fact the choice for the scaling of $N$ with
respect to $\epsilon$, \eqref{Nvs.epsilon-Ext}, was made requiring,
similarly to usual near-horizon decoupling limits \emph{e.g.} see
\cite{AdS-CFT, MAGOO}, the $6d$ string length squared $1/2\pi
T^{(6)}_s$ to scale as $\epsilon$.\footnote{Although the arguments
of this section are mainly made having the far from BPS case in
mind, for the near-BPS case one may give a similar $6d$ description.
The $AdS_3\times S^3$ geometry obtained as near-BPS, near-horizon
limit is a solution to a $6d$ SUGRA in which there is no potential
term for the scalar $X$, \emph{e.g.} see \cite{de-Boer-6d}. In this
case the $AdS_3\times S^3$ solution is the near-horizon limit of
electrically and magnetically charged $6d$ strings. From the $10d$
IIB viewpoint, however, this corresponds to intersecting giants,
which have two of their directions on the $4d$ compact part ${\cal
C}_4$ \cite{Balasubramanian:2007bs}. From this one may read the
effective $6d$ string tension. Recalling the form of metric after
the decoupling limit \eqref{metric-BPS-limit} we have
\[
T^{(6)}_s|_{Near\ BPS}\sim T_3 \epsilon L^2 \sim N\epsilon/L^2\ .
\]

In the limit we are taking, $l_s^2\sim N^{-1/2}\sim \epsilon\to 0$
and hence the $6d$ and $10d$ string scales, as expected, are going
to zero in the same way. This could be used as an alternative way
to argue for the choice made in \eqref{Nvs.epsilon-BPS-preferred}.
This gives a uniform picture for both of the near-BPS and
far from BPS limits, that in both of the cases we scale the
corresponding $6d$ string scale squared as $\epsilon$ to zero
while keeping $L$ fixed and that, certain massless fluctuations of
these $6d$ strings are the degrees of freedom of the $2d$ dual
CFT.\label{footnote-BPS-scaling}}

Assuming that each string carries one unit of electric or magnetic
charge, \footnote{Recalling the form of our $6d$ action
\eqref{6d-action-section3.3} and that  its vacuum solution is an
$AdS_6$ of radius $\sqrt{2/5} L$ with $X=1$, the tension of
electrically and magnetically charged strings are equal.} the number
of electrically or magnetically charged strings are hence
\be\label{6d-charge}%
N_e={2\pi T^{(6)}_s} Q_e={N}\ \frac{\tilde q_2}{L^2}, \qquad N_m={2\pi T^{(6)}_s} Q_m={N}\ \frac{\tilde q_3}{L^2}, %
\ee%
which as expected is exactly the same number as the intersecting
three-brane giants (\emph{cf.} \eqref{Ni-Ji}). As discussed
earlier, the intersecting giant graviton system is not
supersymmetric even at the extremal point. In the reduced $6d$
gravity theory, however, we expect the system of $N_2$
electrically and $N_3$ magnetically charged strings to form a
``dyonic'' $(Q_e,Q_m)$-type string state. The mass (squared) of
this state, as usual BPS dyonic states, is sum of the mass squares
of electrically and magnetically charged strings (see
\eqref{dyon-mass}). Number of this dynoic strings is then the
largest-common-divisor of $N_2$ and $N_3$. As $N_2$ and $N_3$ are
both scaling like $N$, the number of the dyonic strings bound
states formed out of  these strings is then expected to be scaling
like $N$. These $6d$ dynoic $(Q_e,Q_m)$-strings is briefly
discussed in the discussion section and more thorough analysis
will be presented in \cite{progress-6d}.

 As the final step in the entropy function prescription, the
entropy of the black string system (and hence that of the BTZ$\times
S^3$ solution) is given by the value of the entropy function
\eqref{F-plugged} at its minimum, that is it is value at
\eqref{ent-func-solution}, which is
\begin{equation}\label{entropy-ent-func}
    F(Q_e=\tilde q_2, Q_m=\tilde q_3)=\frac{\pi^3\rho_0\epsilon
    }{G_N^{(6)}}\cdot \frac{(v_1v_2)^\frac32}{v_1}.
\end{equation}
Recalling \eqref{ent-func-solution} and that $R^2_{AdS}=v_1,\
R^2_S=v_2$, we see that $F(\tilde q_2,\tilde q_3)=S_{BH}$ given in
\eqref{BH-entropy-5d-Ext} which is in turn equal to
\eqref{3d-entropy}.

\section{Perturbative Addition of The Third
Charge}\label{3rd-charge-section}

So far we have discussed two near-horizon far from BPS
decoupling limits of the two-charge $5d$ black hole type
solutions. It is, however, possible to turn on the third charge.
In this case, again there are two possibilities for an extremal
solution, the BPS case for which $\mu=0$, and the  $\mu=\mu_c$
case where the function $f$ in the metric \eqref{Hif-10d} has a
double horizon at $r=r_h\neq 0$ \cite{Behrndt:1998jd}. In the
generic three-charge extremal but non-BPS case, unlike the
two-charge case, the horizon radius is non-zero and as discussed
in \cite{Morales:2006gm} taking the near-horizon limit leads to
$AdS_2\times S^3$ geometry with unequal $AdS_2$ and $S^3$ radii
\cite{Morales:2006gm}, see also \cite{Astefanesei:2007vh}.

On the other hand, one may ask whether it is possible to add the
third charge as a ``perturbation'' to the two-charge system, that is
adding the third charge $q_1$ such that in the near-horizon scaling
$q_1\ll q_2,q_3$. If this is possible then one expects to find a
rotating BTZ black hole. In this section we show that indeed such a
possibility can be realized for both of the near-BPS and
far from BPS, but non-BPS decoupling limits, respectively discussed
in sections \ref{NH-BPS} and \ref{non-BPS-extremal-limit}. We also
extend black hole entropy  arguments of section
\ref{BH-entropy-section} to these cases.

\subsection{The near-horizon limit: the near-BPS case}\label{rotating-BPS-limit-section}

In this section we extend the limit defined in
\eqref{the-limit-BPS-1}, \eqref{the-limit-BPS-2} by turning on the
third charge $q_1$ in a perturbative manner,
\emph{i.e.}%
\be\label{q1-BPS-limit}%
q_1= \epsilon^2 \hat q_1%
\ee%
while keeping $\hat q_1$ fixed. The rest of parameters are scaled
 the same as before. Let us first consider the case
corresponding to $\mu_1\sim \mu_1^0\neq 1$. In this case
(\emph{cf.}
\eqref{Hif-10d})%
\bse\label{Hif-10d-q1-limit-2}
\begin{align}
 H_1 = 1 + \frac{\hat q_1}{\tilde\rho^2}, &\qquad a_1 = -\frac{1}{L} \sqrt{\hat q_1({\hat\mu}+{\hat q_1})}
\ \frac{1}{\tilde\rho^2+\hat q_1}, \\
H_i = \epsilon^{-1} \frac{\hat q_i}{\tilde \rho^2}\ \ \ i=2,3,
&\qquad
\Delta=\epsilon^{-2} \frac{\hat q_2\hat q_3}{\tilde\rho^4}\ (\mu^0_1)^2\\
f=1-\frac{M}{\tilde \rho^2}&+\frac{J^2}{4\tilde\rho^4}
\end{align}\ese
where $M$ is defined in \eqref{the-limit-BPS-2}, $\hat\mu_c=\hat
q_2\hat q_3/L^2$ and $J^2=4\hat q_1\hat q_2\hat q_3/L^2=4\hat\mu_c\hat
q_1$.

If we redefine $\hat \rho$ as \cite{Myers:2001aq}%
\be\label{rho-shift}%
\rho^2=\tilde\rho^2+\hat q_1%
\ee%
the metric takes the form%
\be\label{NH-BPS-rotating-2}%
\begin{split}
ds^2/\epsilon&=\mu_1^0\left[-\frac{\rho^2 F(\rho)}{R_S^2}dt^2+R^2_S
\frac{d\rho^2}{\rho^2 F(\rho)}+\frac{L^2}{R^2_S}\rho^2 (d\phi_1+a_1
dt)^2+R_S^2 d\Omega_3^2\right]\cr &+
\frac{L^2}{R^2_S}\frac{1}{\mu_1^0}\sum_{i=2,3}\ \hat q_i
\left((d\mu_i/\epsilon)^2+(\mu_i^0)^2 d\psi_i^2\right),
\end{split}
\ee%
where $\psi_i$ are as defined in \eqref{the-limit-BPS-2},
$R_S^4=\hat q_2\hat q_3$ and
\be\label{F-rho}%
F(\rho)=1-\frac{M+2\hat q_1}{\rho^2}+\frac{\hat q_1(\hat \mu+\hat
q_1)}{\rho^4}.%
\ee%
In the above $\hat\mu=M+\hat q_2\hat q_3/L^2$. Note also that  as
discussed in section \ref{NH-BPS} in this case
$d\mu_i/\epsilon=$fixed.

The first line in the  metric \eqref{NH-BPS-rotating-2} describes
an $X_{M,J}\times S^3$ space where $X_{M, J}$ depending on the
values of $M^2$ and $J$ can be a rotating BTZ, conical space or
global $AdS_3$ (see Appendix B for more detailed discussion).
Radius of the $AdS_3$ background (measured in units of
$\sqrt{\epsilon}$) is
\[{\ell}^2=\mu_1^0 R^2_S.\]
The mass $M_{BTZ}$ and angular momentum $J_{BTZ}$ ({\emph{cf.}}
Appendix B) is then~\footnote{The physical angular momentum of the
original $10d$ black-brane (or electric charge of the $5d$ black
hole) corresponding to $q_1$ charge, $J_1$, is related to
$J_{BTZ}$ as
\[ J_1=\frac{N^2\epsilon^2}{4}\ \frac{\mu_c}{L^2} J_{BTZ}.\]
We will comment on this relation in section 6.2.1, in
\eqref{2d-4d}.}
\be\label{M-J-NH-BPS-2}%
M_{BTZ}=\frac{M+2\hat q_1}{\hat \mu_c}=\frac{\hat\mu+2\hat
q_1}{\hat\mu_c}-1,\qquad J_{BTZ}=
2 \sqrt{\frac{\hat q_1 (\hat\mu+\hat q_1)}{\hat\mu_c^2}}.%
\ee%
We should stress that the above metric is a rotating black hole
only when the extremality bound $M_{BTZ}\geq J_{BTZ}$ holds (and
also $\phi\in [0,2\pi]$). In terms of the parameters we have
in our solution, that is%
\be\label{Extremality-Bound-BPS-case}%
M^2\geq 4\hat q_1\hat q_2\hat q_3/L^2.%
\ee %
Note that $M$ can be positive or negative. The above solution is a
black hole at (Hawking) temperature (measured in units of
$\sqrt{\epsilon}\ell$)%
\be\label{rotating-BTZ-Temp-BPS}\begin{split}%
T_{BTZ} =\frac{\sqrt{M^2-4\hat q_1 \hat q_2\hat
q_3/L^4}}{2\pi\rho_h\hat\mu_c},\qquad%
 \rho^2_h= \frac{1}{2\hat \mu_c}\left(M+2\hat q_1+ \sqrt{M^2-4\hat q_1 \hat
q_2\hat
q_3/L^4}\right).\end{split}%
\ee%
For the special case of $M^2=4\hat q_1\hat q_2\hat q_3/L^2$ we have
an extremal rotating BTZ which has $T_{BTZ}=0$.

When $M_{BTZ}\leq -J_{BTZ}\leq 0$, we will have a sensible conical
singularity (see Appendix B) only if
\eqref{Extremality-Bound-BPS-case} holds while $M+2\hat q_1\leq
0$, that is%
\be%
M \leq -2\ Max(\hat q_1, \sqrt{\hat q_1\hat q_2\hat q_3/L^2}),%
\ee%
and if $\gamma$, $\gamma^2\equiv J_{BTZ}-M_{BTZ}$, is a rational
number.

In sum, in order to have a sensible string theory description we
should have%
\be\label{sensibility-BPS}%
M_{BTZ}-J_{BTZ}+1\geq 0,\ %
\ee%
and if $ 0\leq J_{BTZ}-M_{BTZ}\equiv \gamma^2\leq 1$, $\gamma$
should be rational.

In a similar manner one can extend \eqref{the-limit-BPS-1} to the
case with non-zero $\hat q_1$. The result is the same as
\eqref{theta1=0} but the $AdS_3$ part is again replaced with a
rotating BTZ.

It is straightforward to generalize the discussions of section
\ref{BH-entropy-section} to this case and compute the
Bekenstein-Hawking entropy of the above $5d$ black hole solutions
after taking the limit \footnote{Note that the expression for
$\rho_h^2$ is nothing but
\[\rho_h^2=\frac12\left(M_{BTZ}+\sqrt{M_{BTZ}^2-J_{BTZ}^2}\right)\]
yielding
\[\rho_h=\frac12\left(\sqrt{M_{BTZ}-J_{BTZ}}+\sqrt{M_{BTZ}+J_{BTZ}}\right).\]
Therefore, we have a matching between the entropy of the $5d$
blackhole and that of $3d$ rotating BTZ.}
\be\label{BTZ-rotating-entropy}%
S_{BH}=\pi N^2\epsilon^2 \frac{\hat{\mu}_c}{L^2} \rho_h. \ee%
 As we see the entropy,  similarly to the static case of section
\ref{BH-entropy-5d}, scales as $N^2\epsilon^2$. The curvature
radii square, however, scale as $\epsilon\ell^2$. In order for the
gravity description to be valid, one needs to ensure $\epsilon
\ell^2\gg \alpha'$ or $\epsilon \sqrt{N}\gg \ell^2/L^2$, which is
fulfilled by taking $L$ to be parameterically larger than $\ell$,
and scaling $l_s^2\sim \epsilon$ or equivalently $N\sim
\epsilon^{-2}$. In this case, the entropy scales as
$N\sim\epsilon^{-2}\to\infty$.

\subsection{The near-horizon limit: the far from BPS
case}\label{rotating-Ext-section}

In this part we extend the limit discussed in section
\ref{non-BPS-extremal-limit} in equation \eqref{the-limit-static} by
turning on the third charge $q_1$ ``perturbatively'', with the
scaling%
\be\label{q1-extrema-nonBPS-limit}%
 q_1=\epsilon^4\hat q_1\ .%
\ee%
In this case, $H_1\simeq 1$ and $\Delta$ is  given in
\eqref{f-Delta-limit} but now $f$ is %
\be\label{f-rotating-limit}%
f=f_0-\frac{M}{\tilde \rho^2}+ \frac{J^2}{4\tilde\rho^4}, \qquad J^2\equiv 4\frac{{\hat q}_1\,q_2\,q_3}{L^2}=4\mu_c\hat q_1%
\ee%
where $f_0=1+\frac{q_2+q_3}{L^2}$. After taking the above limit the metric takes the form
\be\label{rotating-extremal-limit}%
ds^2=\mu_1\left[ R^2_{AdS}\ ds^2_{BTZ}+R^2_S\ d\Omega^2_3\right]+ \frac{1}{\mu_1} d{\cal M}_4^2%
\ee%
where $d{\cal M}_4^2$ is given in \eqref{M4-metric}, $R^4_S=q_2q_3, \ R^2_{AdS}=R^2_S/f_0$ and
\be\label{BTZ-rotating}%
ds^2_{BTZ}=-N(\rho) d\tau^2+\frac{d\rho^2}{N(\rho)}+\rho^2
(d\varphi-N_{\varphi}d\tau)^2%
\ee%
in which%
\be%
 N(\rho)=\rho^2- M_{BTZ}+\frac{J_{BTZ}^2}{4\rho^2},\qquad
N_{\varphi}=\frac{J_{BTZ}}{2\rho^2} \ ,%
\ee%
with%
\be\label{M-J-Extremal}%
 M_{BTZ}=\frac{M}{\mu_c},\qquad%
 J_{BTZ}=2\sqrt{\frac{f_0\hat q_1}{\mu_c}}\ .
\ee%
Note that although the metric \eqref{BTZ-rotating} has the standard
form of a  rotating BTZ black hole \cite{BTZ} (see also Appendix B)
we should stress that the range of the angular coordinate $\varphi$
is $[0,2\pi\epsilon]$.
 The new time coordinates $\tau,\rho$ are related to
the original $10d$ coordinates as in \eqref{rho-tau-t-r}. The
above geometry has the interpretation of rotating BTZ
only if $N(\rho)=0$ has real solutions, that is when %
\be\label{Extremality-bound-Ext.}
M^2\geq 4\mu_c f_0 \hat q_1.%
\ee%

The Bekenstein-Hawking entropy of the above rotating BTZ is%
\be\label{entropy-rot-BTZ-Ext.}%
S_{BH}=\pi\rho_h\ \frac{1}{\sqrt{f_0}}\frac{\mu_c}{L^2}\
N^2\epsilon\ ,\quad \rho_h=\frac12\left(\sqrt{M_{BTZ}+J_{BTZ}}+\sqrt{M_{BTZ}-J_{BTZ}}\right), %
\ee%
which turns out to be exactly the expression one would obtain
after taking the decoupling limit on the Bekenstein-Hawking
entropy of the corresponding three-charge $5d$ black hole.

It is also immediate to see that repeating the entropy function
analysis of section \ref{6d-analysis-section} for the rotating BTZ
case, we will exactly obtain the entropy for the system of
corresponding  $6d$ black strings. The $6d$ analysis for this case
is again suggestive of existence of $6d$ string picture. We will
comment further on this point in section
\ref{2d-CFT-NExt-section}.

\section{The Dual Gauge Theory
Descriptions}\label{dual-gauge-section}

So far we have shown that one can take specific near-horizon,
near-extremal limits over $10d$ type IIB solutions which are
asymptotically $AdS_5$. As such one would expect that these
solutions, the limiting procedure and the resulting geometry after
the limit should have a dual description via $AdS_5/CFT_4$. On the
other hand, after the limit we obtain a space which contains
$AdS_3\times S^3$ and hence there should also be another dual
description in terms of a $2d$ CFT. In this section we intend to
study both $4d$ and $2d$ dual CFT descriptions.

\subsection{Description in terms of $d=4, {\cal N}=4$
SYM}\label{SYM-subsection}

In this section we first identify the operators of ${\cal N}=4,\
d=4$ $U(N)$ SYM theory, by specifying their $SO(4,2)\times SO(6)$
quantum numbers, corresponding to the two-charge gravity solutions
discussed in earlier sections. We then translate  taking the
near-horizon, near-extremal limits in the dual ${\cal N}=4$ theory
and identify the sector of the gauge theory operators
corresponding to string theory on the decoupled backgrounds for
both of the near-BPS and far from BPS cases.

According to the standard AdS/CFT dictionary \cite{AdS-CFT}, the
scaling dimension of operators $\Delta$ and their $R$-charge $J_i$
respectively correspond to the ADM mass and angular momentum of the
objects in the gravity side. Explicitly, for the two-charge case of
our interest, the operators are specified by three quantum numbers
\cite{Myers:2001aq,Balasubramanian:2007bs,Gubser:2004xx}:
\footnote{In our conventions  the BPS condition in the gauge theory
side is written as $\Delta=\sum_i J_i$.} %
\be\label{Delta,J}
\begin{split}%
\Delta &=L\cdot M_{ADM}=\frac{N^2}{2L^2}\ (\frac{3}{2}\mu+q_2+q_3)\
,\cr J_i &=\frac{\pi L}{4G_5} \tilde q_i=\frac{N^2}{2}\ \frac{\tilde
q_i}{L^2}\ .
\end{split}\ee%
where $M_{ADM}$ has the same expression as $M$ (\ref{5d-ADM-mass-charge}b) but the last term, the Casimir energy, has been
dropped. Operators in this sector are singlets of the $SO(4)\in SO(4,2)$.
As we see in both cases, if $\mu$ and $q_i$ are finite, $\Delta$ and $J_i$ both scale like $N^2$.

In both of the near-BPS and far from BPS limits we
are taking the 't Hooft coupling, $\lambda=L^4/l_s^4$ to infinity
and one should not expect the dual $4d$ field theory to give a
useful, \emph{i.e.}  a perturbative, description. On the other hand,
after  BMN \cite{BMN}, we have learned that the effective expansion
parameters of the $4d$ gauge theory may be different in sectors of
large $R$-charges such that for specific ``almost-BPS'' operators
the effective (or ``dressed'') 't Hooft coupling and the genus
expansion parameter remains finite. In this subsection we try to
extend the ideology of BMN to the new ``almost-BPS'' as well as
``almost-extremal'' sectors.

\subsubsection{Near-horizon near-BPS limit, ${\cal N}=4$ SYM
description}\label{NBPS-SYM-section}

In the near-BPS limit case together with some of the coordinates we
also scale $\mu$ and $q_i$ as $\epsilon$. As discussed in section
\ref{BH-entropy-BPS-section} (see also footnote
\ref{footnote-BPS-scaling})  we need to also scale $N\sim
\epsilon^{-2}$. Therefore, in this limit  $\Delta$ and $J_i$ of the
operators scale as:%
\be\label{Delta-J-BPS-limit}
\begin{split}
\Delta &=\frac{N^2\epsilon}{2}\ (\hat q_2+\hat q_3+{\cal
O}(\epsilon))/L^2\sim N^{3/2}\to\infty\\ J_i&=
\frac{N^2\epsilon}{2}(\hat q_i+ {\cal O}(\epsilon))/L^2\sim N^{3/2}\
.\end{split}
\ee%
That is, the sector of the ${\cal N}=4$ $U(N)$ SYM operators
corresponding to the geometries in question have large scaling
dimension and $R$-charge, $\Delta\sim J_i\sim N^{3/2}$. In the
same spirit as the BMN limit \cite{BMN}, one can find certain
combinations of $\Delta$ and $J_i$ which are finite and describe
physics of the operators after the limit. To find these
combinations we recall the way the limit was taken, \emph{i.e.}
\eqref{the-limit-BPS-1}, \footnote{In the near-BPS case we
discussed two limits, $\mu_1\simeq 1$ and $\mu_1\neq 1$ of
\eqref{the-limit-BPS-2}. For the latter one also scales $\psi_i$
with respect to $\phi_i$. As a result, (\ref{Delta-J-BPS-1}a)
remains unchanged while (\ref{Delta-J-BPS-1}b) is modified to
$-i\frac{\partial}{\partial\psi_i}=\epsilon^{1/2} J_i$. In this
section, however, we are only going to utilize
(\ref{Delta-J-BPS-1}a).} and in particular note that \footnote{In
the Penrose-BMN limit, {\it a la} Tseytlin, the $AdS_5\times S^5$
coordinates $\tau,\psi$ are related to the ``light-cone''
coordinates $x^+,x^-$ as
\cite{Tseytlin}%
\[
t=x^+,\qquad  \psi=x^+-\frac{1}{R^2} x^-%
\]
where $R$ is the $AdS$ radius which is taken to infinity as $N^{1/4}$.
}%
\begin{subequations}\label{Delta-J-BPS-1}%
\begin{align}%
iL\frac{\partial}{\partial \tau}&=iL\frac{\partial}{\partial
t}+i\sum_{i=2,3} \frac{\partial}{\partial \phi_i} =\Delta-\sum_{i=2,3} J_i\\
-i\frac{\partial}{\partial \psi_i}&=-i\frac{\partial}{\partial
\phi_i}=J_i
\end{align}%
\end{subequations}%
Up to leading order we have
\be\label{Delta-J-BPS-2}%
\begin{split}%
\Delta-\sum_{i=2,3}
J_i&=\frac{N^2\epsilon^2}{4}\frac{\hat\mu}{L^2}\ , \cr
J_i&=\frac{N^2\epsilon}{2}\frac{\hat q_i}{L^2}\ .
\end{split}%
\ee%
As we see $\Delta-\sum J_i$ scales as $N^2\cdot
N^{-1}=N\to\infty$, while $J_i\sim N^{3/2}$ and therefore the
``BPS deviation parameter'' \cite{Berenstein03}
\be\label{BPS-deviation} %
\eta_i\equiv \frac{\Delta-\sum_i
J_i}{J_i}\sim \epsilon\sim N^{-1/2}\to 0\ , %
\ee%
and hence we are dealing with an ``almost-BPS''
sector.\footnote{It is instructive to make parallels with the BMN
sector \cite{BMN}. In the BMN sector we are dealing with operators
with
\[
\Delta\sim J\sim N^{1/2},\qquad {\rm while}\qquad \Delta-J\ =finite,
\]
implying that, similarly to our case, $\eta_{BMN}\sim N^{-1/2}\to 0$.}
Moreover,
$\Delta-\sum J_i$ is linearly proportional to non-extremality
parameter $\hat\mu$. It is also notable that $S_{BH}$
\eqref{entropy-near-BPS} scales the same as $\Delta-\sum J_i$.

To write \eqref{Delta-J-BPS-2} in terms of the gauge theory
parameters we need to replace $\epsilon$ for  parameters of the
gauge theory, which we choose%
 \be\label{eps-lambda}
\epsilon=\frac{2}{\sqrt N}\ .
\ee%
In sum, the sector we are dealing with is  composed of ``almost (at
most) 1/4 BPS'' operators of  $U(N)$ SYM with \footnote{Note that
the value for $\Delta-\sum J_i$ read from the gravity side, via
AdS/CFT, corresponds to the strong 't Hooft coupling regime of the
gauge theory.}%
\be\label{SYM-observables}%
\begin{split}
\Delta\sim J_i\sim N^{3/2} &,\qquad \lambda=g^2_{YM}N \sim N\to \infty \\
\frac{J_i}{N^{3/2}}\equiv \frac{{\hat q}_i}{L^2}=fixed &, \qquad
(\Delta-\sum_{i=2,3} J_i)\cdot \frac{1}{N}=\frac{\hat
\mu}{L^2}=fixed.
\end{split}%
\ee%
The  dimensionless physical quantities that describe this sector are
therefore $\hat q_i/L^2,\ \hat \mu/L^2$ and $g_{YM}$.

To specify the sector completely we should also determine the
\emph{basis} we use to contract the $N\times N$ $U(N)$ gauge
indices. This could be done by giving the (approximate) shape of
the corresponding Young tableaux. To this end we recall the
interpretation of the original $10d$ geometry in terms of the
back-reaction of the intersecting giant gravitons and that giant
gravitons and their open string fluctuations are described by
(sub)determinant operators \cite{sub-det, open-string, deMello,
Berenstein03}. Here we are dealing with a system of intersecting
multi giants. The ``number of giants'' in each stack in the near-BPS, near-horizon limit is \eqref{Ni-Ji}%
\be\label{giant-number-BPS}%
 N_i=N\epsilon\cdot \frac{\hat q_i}{L^2}=2N^{1/2}\ \frac{\hat
 q_i}{L^2}\ ,
\ee%
and therefore, $\Delta-\sum_i J_i=\frac{N_2N_3}{4} \frac{\hat
\mu}{\hat \mu_c}.$

Finally, let us consider the rotating case of section
\ref{rotating-BPS-limit-section}, where besides $J_2,\ J_3$ we have
also turned on the third $R$-charge $J_1$,
\be%
J_1=\frac{N^2\epsilon^2}{2}\cdot\frac{1}{L^2} \sqrt{\hat q_1(\hat
q_1+\hat\mu)}\ .
\ee%
As we see $\Delta-\sum_{i=2,3} J_i\sim J_1\sim N^2\epsilon^2\sim
N\to \infty$. (Note that in this case, as we have also turned on
the third charge $q_1$, $\Delta$ is not the expression given in
\eqref{Delta,J} and one should use \eqref{5d-ADM-mass-charge}.)
Instead of $\Delta-\sum_{i=2,3} J_i$ it is more
appropriate to define another positive definite quantity:%
\be\label{Delta-J-3charge-BPS}%
 \Delta-\sum_{i=1}^3
J_i=N \cdot \left(\frac{\hat\mu+2\hat q_1- \sqrt{(\hat\mu
+2\hat q_1)^2-\hat\mu^2}}{L^2}\right)\ \geq 0\ . %
\ee%
It is remarkable that the above BPS bound is exactly the same as the
bound \eqref{sensibility-BPS}. This bound is more general than just
the extremality bound of the rotating BTZ black hole $M_{BTZ}\geq
J_{BTZ} \geq 0$. This bound besides the rotating black hole cases
also includes the case in which we have a conical singularity which
could be resolved in string theory (\emph{cf.} Appendix B and
section 5.1). We will comment on this point further in section
\ref{2d-CFT-NBPS-section}.

\subsubsection{Near-horizon far from BPS limit, ${\cal N}=4$ SYM
description}\label{NExt.-SYM-section}

Since in the near-horizon, far from BPS limit of
\eqref{the-limit-static} we do not scale $\mu$ and $q_i$'s, in
this case we expect to deal with a sector of ${\cal N}=4$ SYM in
which $\Delta\sim J_i \sim N^2$ and, as discussed in
\ref{BH-entropy-nonBPS-section}, $N\sim \epsilon^{-1}$. To deduce
the correct ``BMN-type'' combination of $\Delta$ and $J_i$ which
correspond to physical observables, we again recall the way the
limit has been taken, and in particular \be \tau={\epsilon}\
\frac{R_S}{R_{AdS_3}}\frac{t}{L},\qquad \phi_i = \psi_i +
\frac{{\tilde q}_i R_{AdS_3}}{q_i  R_S } \frac{\tau}{\epsilon},\
i=2,3\ . \ee Therefore, $-i\frac{\partial}{\partial
\psi_i}=-i\frac{\partial}{\partial \phi_i}=J_i$ and
\be\label{Delta-J-extremal}%
\begin{split}%
{\cal E}\equiv -i\frac{\partial}{\partial
\tau}=-\frac{R_{AdS_3}}{\epsilon \ R_S}\left(i
L\frac{\partial}{\partial t}+ i\sum_{i=2,3} \frac{\tilde q_i}{q_i}
\frac{\partial}{\partial \phi_i}\right) =
-\frac{R_{AdS_3}}{\epsilon \
R_S}\left(\Delta-\frac{2L^2}{N^2}\sum_{i=2,3}
\frac{J_i^2}{q_i}\right)
\end{split}%
\ee%
The last equality  can be understood in an intuitive way. In the
near-extremal case we are indeed dealing with massive giant
gravitons which are far from being BPS and hence are behaving like
\emph{non-relativistic} objects which are rotating with angular
momentum $J_i$ over circles with radii $R_i$,
$R_i^2=\frac{L^2}{R^2_S} q_i$ \eqref{M4-metric}. Therefore, the
kinetic energy of this rotating branes is proportional to $\sum
J^2_i/q_i$. As discussed in section \ref{3d-entropy-NExt-section}
in our limit $\epsilon\sim 1/N$ which for convenience we choose
\be\label{Nvs.epsilon-brane-mass}%
\epsilon=\frac{4}{N}.%
\ee%

 Recalling that $\Delta$ is measuring the ``total'' energy of
the system, then ${\cal E}$ should have two parts: the rest mass
of the system of giants and the energy corresponding to the
``internal'' excitations of the branes. To see this explicitly we
recall \eqref{Delta,J}, \eqref{5d-ADM-mass-charge} and
\eqref{6d-string-tension}, yielding
\be\label{Delta-J-extremal-2}%
\begin{split}
{\cal E}&=\frac{R_{AdS_3}}{R_S}\cdot{\frac{N^2}{4\epsilon}}\cdot
\frac{\mu}{L^2}\cr &= {\cal E}_0+ \frac{R_{AdS_3}}{R_S}\cdot (2\pi
T^{(6)}_s M)
\end{split}%
\ee%
 where have used
$\mu=\mu_c+\epsilon^2 M$ ($M$ is related to the mass of BTZ black
hole \eqref{Rs-RAdS}),
and%
\be\label{E0}%
{\cal E}_0=\frac{R_{AdS_3}R_S^3}{16L^4}\cdot N^3.
\ee%
${\cal E}_0$  which is basically ${\cal E}$ evaluated at
$\mu=\mu_c$, is the rest mass of the brane system.\footnote{One
should keep in mind that at the extremal point the system is not BPS
and hence the ``rest mass'' of the giants system is not simply sum
of the masses of individual stacks of giants and already contains
their ``binding energy'' (stored in the deformation of the giant
shape from the spherical shape). Nonetheless, it should still be
proportional to the number of giants times mass of a single giant.
In the $6d$ language, as suggested in section
\ref{6d-analysis-section}, this corresponds to formation of a $6d$
$(Q_e,Q_m)$-string. Eq.\eqref{E0}, however, seems to suggest a
simpler interpretation in terms of dual giants \cite{dual-giant}.
Inspired by the expression for the $10d$ five-form flux and
recalling that the IIB five-form is self-dual, the system of giants
we start with, \emph{e.g.} through SUGRA solution
\eqref{10-dim-general}, may also be interpreted  as spherical
three-branes wrapping  $S^3\in AdS_5$ while rotating on $S^5$.   After the limit, we are dealing with a system of dual giants wrapping the $S^3\in AdS_3\times S^3$, of
radius $R_S$. The mass of a single such dual giant $m_0$ (as
measured in $R_{AdS_3}$ units and also noting the scaling of $AdS_5$
time with respect to $AdS_3$ time) is then
\[ \frac{m_0}{R_{AdS_3}/\epsilon}=T_3(2\pi^2 R_S^3)= \frac{R^3_S}{L^4}\cdot N.
\]
The number of dual giants is again proportional to $N$ and hence
one expects the total ``rest mass'' of the system $m_0$
to be proportional to $N^3 R^3_S$.}%

${\cal E}-{\cal E}_0$ which is proportional to $T^{(6)}_s M$
corresponds to  the fluctuations of the giants about the extremal
point. The fact that ${\cal E}-{\cal E}_0$ is proportional to
$T^{(6)}_s M$ indicates that it can be recognized as a
fluctuations of a $6d$ string. Recall also that from the $10d$
viewpoint, the $6d$ strings are uplifted to three-brane giants
with two legs along the ${\cal M}_4$ directions. Therefore, ${\cal
E}-{\cal E}_0$ corresponds to (three) brane-type fluctuations of
the original ``intersecting giants''.

In sum, from the $U(N)$ SYM theory viewpoint the sector describing
the near-horizon far from BPS limit consists of operators
specified with%
 \be\label{Ext-SYM-sector}
 \begin{split}
 \Delta\sim J_i\sim N^2, &\qquad \lambda\sim N\to \infty,\cr
 \frac{J_i}{N^2}\equiv \frac{\tilde q_i}{2L^2}=fixed, &\qquad \frac{{\cal E}-{\cal
 E}_0}{N}=fixed\ ,
 \end{split}
 \ee%
where ${\cal E}$, ${\cal E}_0$ in  equations
\eqref{Delta-J-extremal},\eqref{Delta-J-extremal-2} and \eqref{E0}
are defined in terms of $\Delta$, $J_i$.

As discussed in section \ref{rotating-Ext-section} one may obtain a
rotating BTZ if we turn on the third $R$-charge in a perturbative
manner. In the $4d$ gauge theory language this is considering the
operators which besides being in the sector specified by
\eqref{Ext-SYM-sector} carry the third $R$-charge $J_1$,
$J_1\sim N^2\epsilon^2\sim 1$. Explicitly,%
\be\label{3rd-charge-Ext}%
 J_1=\frac{N^2}{2L^2} \epsilon^2\ \sqrt{\hat q_1\mu_c}%
\ee%
One should, however, note that in terms of the $AdS_3$ parameters,
since $\varphi=\epsilon\phi$, then%
\be\label{varphi-momentum}%
{\cal J}\equiv
-i\frac{\partial}{\partial\varphi}=-i\frac{1}{\epsilon}\frac{\partial}{\partial\phi}=
 \frac{J_1}{\epsilon}=\frac{N^2\epsilon}{2}\ \frac{\mu_c}{L^2}\
 \sqrt{\frac{\hat q_1}{\mu_c}}%
 \ee%
As we see ${\cal J}$, similarly to ${\cal E}-{\cal E}_0$, is also
scaling like $N^2\epsilon\sim N$ in our decoupling limit. When
$J_1$ is turned on the expressions for the $\Delta$ and hence
${\cal E}$ are modified, receiving contributions from $q_1$. These
corrections, recalling \eqref{5d-ADM-mass-charge} and that $q_1$
scales as $\epsilon^4$ \eqref{q1-extrema-nonBPS-limit}, vanish in
the leading order. However, one may still define physically
interesting combinations like ${\cal E}-{\cal E}_0\pm {\cal J}$.
We will elaborate further on this point in section
\ref{2d-CFT-NExt-section}.

Before closing this subsection some comments are in order:
\begin{itemize}
\item{ The remarkable point which is seen directly from
\eqref{Delta-J-extremal} is that
 $-{\cal E}$ is negative definite, \emph{i.e.} there is an \emph{extremality bound}:
\be\label{Ext-bound} \Delta- \sum_i f_i(J_i)\leq 0 .
\ee%
where
\[
f_i(J_i)= \frac{2L^2}{N^2} \frac{J_i^2}{q_i}. \]%
 (Note that one can
express $q_i$ in terms of the $J_i$'s but since the explicit
expressions are not illuminating we do not present them here.)
 This could be thought of as a complement to the usual BPS bound, $\Delta-\sum_i J_i\geq
 0$.}
\item{
 We note that both ${\cal E}-{\cal E}_0$ and ${\cal J}$
scale as $N^2\epsilon\sim N$ which is the same scaling as the
black hole entropy \eqref{BH-entropy-5d-Ext}.}

\item{ Finally, the system of original intersecting giants is composed of two
stacks of D3 giants each containing $N_i=N\frac{\tilde q_i}{L^2}$
branes and $N_i\sim N\to\infty$.}
\end{itemize}

\subsection{Description in terms of the dual $2d$ theory}\label{2d-descriptions-subsection}

As we showed in either of the near-BPS or far from BPS
near-horizon limits we obtain a spacetime which has an
$AdS_3\times S^3$ factor. This, within the AdS/CFT ideology, is
suggesting that (type IIB) string theory on the corresponding
geometries should have a dual $2d$ CFT description. In this
section we elaborate on this $2d$ description.

\subsubsection{Near-BPS case, the dual $2d$ CFT description}\label{2d-CFT-NBPS-section}

This case was discussed in \cite{Balasubramanian:2007bs} and
references therein and hence we will be  brief about it. The metric
in this case takes  the same form as the near-horizon limit of a
D1-D5 system, though the $AdS_3$ is obtained to be in \emph{global}
coordinates. This could be understood noting that the geometry
\ref{10-dim-general}, in the two-charge case, corresponds to a
system of \emph{smeared} giant D3-branes intersecting on a circle.
In the near-horizon limit, however, we take the radius of the giants
to be very large (or equivalently focus on a very small region on
the worldvolume of the spherical brane) while keeping the radius of
the intersection circle to be finite (in string units). Therefore,
upon two T-dualities on the D3-branes along the ${\cal C}_4$
directions the system goes over to a D1-D5 system but now the D1 and
D5 are lying on the circle (D5 has its other four directions along
${\cal C}_4$). The situation is essentially the same as the usual
D1-D5 system, \emph{e.g.} see \cite{MAGOO, David:2002wn} for
reviews, with only an important difference
\cite{Balasubramanian:2007bs}. Here we just give the dictionary from
our conventions and notations of the usual D1-D5 system (for a
detailed review see \cite{David:2002wn}) and those of
\cite{Maldacena:2000dr, strominger}, and discuss the difference.
\begin{itemize}
\item{ Number of D-strings $Q_1$ and number of D5 $Q_5$ are respectively equal to the number of giants in each stack
$N_2$ and $N_3$ \eqref{giant-number-BPS}.}
\item{
The degrees of freedom are coming from four DN modes of open strings
stretched between intersecting giants which are in $(N_2,\bar N_3)$
representation of $U(N_2)\times U(N_3)$.}\footnote{It has been
shown, from DBI action analysis \cite{heghog} and using the
description of giants in the ${\cal N}=4$ SYM in \cite{vijay-etal},
that similarly to flat D-brane case, when we have a $N$ number of
giants sitting on top of each other the low energy effective field
theory becomes a $U(N)$ gauge theory on the giant.}
\item{
In taking the near-horizon, near-BPS limit we are focusing on a
narrow strip on $\mu_2, \mu_3$ directions and hence our BTZ$\times
S^3\times {\cal C}_4$ geometry and in this sense the corresponding ${\cal N} = ( 4,4)$
$2d$ CFT description is only describing the narrow strips on the
original $5d$ black hole. Therefore, our $5d$ black hole is
described in terms of not a single $2d$ CFT, but a collection of
(infinitely many of) them. The \emph{only} property which is
different among these $2d$ CFT's is their central charge. The
``metric'' on the space of these $2d$ CFT's is exactly the same as
the metric on ${\cal C}_4$. Therefore, as far as the entropy and
the overall (total) number of degrees of freedom are concerned,
one can define an \emph{effective central charge} of the theory
which is the integral over the central charge of the theory
corresponding to each strip \cite{Balasubramanian:2007bs}. To
compute the central charge we use the Brown-Henneaux central
charge formula \cite{Brown-Henneaux}, \[c=\frac{3\,R_{AdS}}{2\
G^{(3)}}\] and recall that in our case for each strip
$R_{AdS}^2=\mu_1^0 R^2_S$, that $G^{(3)}\propto
\sqrt{\mu_1^0}/\mu^0_2\mu^0_3$ and \emph{effective} total central
charge is obtained by integrating strip-wise $c$ over the ${\cal
C}_4$. Noting that the central
charge of the usual D1-D5 system is given by $6Q_1Q_5$, and that%
\[\int_{\mu_2^2+\mu_3^2\leq 1} \mu_2\mu_3 d\mu_2 d\mu_3=\frac18, \]
the \emph{effective} central charge of the system is
\be\label{central-charge-BPS}%
c_L=c_R=c=3N_2N_3=12 N\cdot\frac{\hat\mu_c}{L^2}.%
\ee%
It is notable that in our case the central charge $c\sim
N\to\infty$.}

\item{In contrast to the results of \cite{Balasubramanian:2007bs} \emph{e.g.} eq.(2.22) or eq.(4.79) there,
we should stress that in our case the entropy, and hence the
central charge $c$ are scaling like $N$, as opposed to $N^2$
there. This difference appears recalling that in our case the
entropy scales as $N^2\epsilon^2$ and that $\epsilon^2\sim 1/N$.}

\item{As discussed in the Appendix B the generic solution can be a
rotating BTZ black hole or conical singularity, if%
\[ M-J\geq -1.\]}
\item{The $2d$ CFT is described by $L_0, \bar L_0$
(respectively equal to the left and right excitation number of the
$2d$ CFT $N_L$ and $N_R$, divided by $N_2N_3$)
 which are related
to the BTZ black hole mass and angular momentum \cite{David:2002wn}
as %
\be\label{2dvs.BTZ}%
L_0=\frac{6}{c}N_L=\frac{1}{4}({M_{BTZ}}-J_{BTZ}), \qquad
\bar L_0=\frac{6}{c}N_R=\frac{1}{4}({M_{BTZ}}+J_{BTZ}).%
 \ee%
The above expressions for $L_0, \bar L_0$ are given for
$M_{BTZ}-J_{BTZ}\geq 0$ when we have a black hole description.
When $-1\leq M_{BTZ}-J_{BTZ} <0$, we need to replace them with
$L_0=-\frac{c}{24} a^2_+$, $\bar L_0=-\frac{c}{24} a_-^2$ (in the
conventions introduced in the Appendix B) \cite{ David:2002wn,
Justin-Mandal}. In the special case of \emph{global} $AdS_3$
background, where $a_+=a_-=1/2$ formally corresponding to
$M_{BTZ}=-1, J_{BTZ}=0$,  the ground state is describing an NSNS
vacuum of the ${\cal N} = ( 4,4)$ $2d$ CFT \cite{strominger, Maldacena:2000dr}.\footnote{ In
\emph{global} $AdS_p$ spaces, when $p$ is odd the expression for
the ADM mass has a Casimir energy \cite{ADM-mass-charge}; for
$AdS_3$, in units of AdS radius ($R$) , the Casimir energy is
given by $R/ 8 G_N^{(3)}$.}
 The expressions for $M_{BTZ}$ and $J_{BTZ}$ in terms of the
system of giants are given in \eqref{M-J-NH-BPS-2}.}
\item{With the above identification, it is readily seen that the Cardy formula for the entropy of
a $2d$ CFT%
\be\label{Cardy-entropy}%
\begin{split}%
S_{2d\ CFT} &= 2\pi\left(\sqrt{cN_L/6}+\sqrt{cN_R/6}\right)\cr%
&= \frac{\pi}{6}\
c\left(\sqrt{M_{BTZ}-J_{BTZ}}+\sqrt{M_{BTZ}+J_{BTZ}}\right)
\end{split}
\ee%
exactly reproduces the expressions for the entropy we got in the
previous section, \eqref{BTZ-rotating-entropy}, once we substitute
for the central charge from \eqref{central-charge-BPS} and
$M_{BTZ},\ J_{BTZ}$ from \eqref{M-J-NH-BPS-2}.}
\item{Although the entropy and the energy of the system (which are both proportional to the
central charge) grow like $N$ and go to infinity in the limit we
are interested in, the temperature and the horizon size
\eqref{rotating-BTZ-Temp-BPS}  remain finite. }
\item{It is also instructive to directly compare the $4d$ description discussed in \ref{NBPS-SYM-section}  and the $2d$
field theory descriptions. Comparing the expressions for $M_{BTZ},
J_{BTZ}$ and $\Delta-\sum_{i=2,3} J_i$, $J_1$, we see that
they match; explicitly%
\be\label{2d-4d}%
\Delta-\sum_{i=2,3} J_i=\frac{c}{12} (M_{BTZ}+1),\qquad
J_1=\frac{c}{12} J_{BTZ}\ . %
\ee%
This is very remarkable because it makes a direct contact between
the $2d$ and $4d$ gauge theory descriptions. The $4d$ gauge theory
BPS bound, \emph{i.e.} $\Delta-\sum_{i=1,2,3}\geq 0$ now
translates into the bound $M_{BTZ}-J_{BTZ}\geq -1$. This means
that the $4d$ gauge theory, besides being able to describe the
rotating BTZ black holes, can describe the conical spaces too. In
other words, $\Delta-\sum_{i=1}^3 J_i=0$ and $N\frac{\hat
\mu_c}{L^2}$ respectively correspond to global $AdS_3$ and
massless BTZ cases and when
\[
0<\Delta-\sum_{i=1}^3 J_i<\frac{c}{12}=N\frac{\hat \mu_c}{L^2}\ ,
\]
the $4d$ gauge theory is describing a conical space, provided that
$\gamma$,
\[
\gamma^2\equiv \frac{12}{c}\left(\Delta-\sum_{i=1}^3 J_i\right)-1,
\]
is a rational number. This is of course expected if the dual gauge
theory description is indeed describing string theory on the
conical space background. One should also keep in mind that
entropy and temperature are sensible only when
$\Delta-\sum_{i=1}^3 J_i\geq \frac{c}{12}$; for smaller values the
degeneracy of the operators in the $4d$ gauge theory is not large
enough to form a horizon of finite size (in $3d$ Planck units).}
\end{itemize}

\subsubsection{Far from BPS  case, the dual  $2d$ CFT description }\label{2d-CFT-NExt-section}

As discussed before, in the near-horizon limit over a
near-extremal two-charge black hole we again obtain an
$AdS_3\times S^3$ in which the $AdS_3$ and $S^3$ factors have
different radii, moreover, although locally $AdS_3$, the
coordinate parameterizing $S^1\in AdS_3$ is ranging over
$[0,2\pi\epsilon]=[0,8\pi/N]$. As such, one expects the dual $2d$
CFT description to have somewhat different properties than the
standard D1-D5 system. Based on the analysis and results of
previous sections we conjecture that there exists a $2d$ CFT which
describes the $6d$ string theory on this $AdS_3\times S^3$
geometry. This string theory could be embedded in the $10d$ IIB
string theory on the background \eqref{rotating-extremal-limit}.

Here we just make some remarks about this conjectured $2d$ CFT and
a full identification and analysis of this theory is postponed for
future works:
\begin{itemize}
\item{ This $2d$ CFT  resides on the $R\times S^1$ causal boundary of
the $AdS_3\times S^3$ geometry ({\emph{cf.}} discussions of
section \ref{non-BPS-extremal-limit}).\footnote{It is worth noting
that in terms of the coordinates $t$ and $\phi_1$ of the original
$AdS_5$ background, as noted in \eqref{boundary-near-Ext} we have
a space which looks like a (supersymmetric) orbifold of $AdS_3$
\cite{Null-orbifolds}, by $Z_{\epsilon^{-1}}$, that is an
$AdS_3/Z_{N/4}$. It is desirable to understand our analysis from
this orbifold viewpoint.}}

\item{Noting \eqref{boundary-near-Ext}, one may use
the Brown-Henneaux analysis \cite{Brown-Henneaux} to compute the
central charge of this $2d$ CFT%
\be\label{central-charge-Extremal}%
c=\frac{3R_{AdS_3}\epsilon}{2 G^{(3)}_N}= 12\frac{\mu_c}{L^2\sqrt{f_0}}\ N\ . %
\ee%
As we see in this case the  expression for the central charge,
except for the $1/\sqrt{f_0}$ factor, is the same as that of the
near-BPS case \eqref{central-charge-BPS}, and scales like
$N\to\infty$ in our limit.}

\item{The $5d$ or $3d$ black hole entropies given in
\eqref{entropy-rot-BTZ-Ext.} take exactly the same form obtained
from counting the number of microstates of a $2d$ CFT, \emph{i.e.}
the Cardy formula \eqref{Cardy-entropy}, with the central charge
\eqref{central-charge-Extremal} and $M_{BTZ}$ and $J_{BTZ}$ given
in \eqref{M-J-Extremal}.}

\item{As discussed in section \ref{NExt.-SYM-section}, there is a
sector of ${\cal N}=4$, $d=4$ SYM which describes the IIB string
theory on the background \eqref{rotating-extremal-limit}. This
sector is characterized by ${\cal E}-{\cal E}_0$ and ${\cal J}$.
From \eqref{Delta-J-extremal-2} and \eqref{varphi-momentum} one
can
readily express the $4d$ parameters in terms of $2d$ parameters, namely:%
\be\label{4d-2d-Extremal}%
{\cal E}-{\cal E}_0=\frac{c}{12} M_{BTZ}\ , \qquad {\cal
J}=\frac{c}{12} J_{BTZ}\ , %
\ee%
where $c$ is given in \eqref{central-charge-Extremal} and
$M_{BTZ}, J_{BTZ}$ are given in \eqref{M-J-Extremal}. The above
relations have of course the standard form of the usual D1-D5
system, and/or the near-BPS case discussed in section
\ref{2d-CFT-NBPS-section}. Note, however, that in this case ${\cal
E}-{\cal E}_0$ is measuring the mass of the BTZ with the zero
point energy set at the massless BTZ case (rather than global
$AdS_3$).}

\item{As discussed in sections \ref{6d-analysis-section} and
\ref{NExt.-SYM-section} we expect the degrees of freedom of this
$2d$ CFT to correspond to $10d$ IIB string states on the
$AdS_3\times S^3$ geometry discussed in
\eqref{6d-action-section3.3}, which in turn  correspond to
brane-like excitations about the extremal intersecting giant
three-branes. It is of course desirable to make this picture precise
and explicitly identify the corresponding $2d$ CFT.}
\end{itemize}
\section{Summary and Discussion}\label{discussion-section}

In this paper we extended the analysis of
\cite{Balasubramanian:2007bs} and discussed in more details the
near-horizon decoupling limits of the near-extremal two-charge
black holes of $U(1)^3$ $d=5$ gauged SUGRA. We showed that there
are \emph{two} such decoupling limits, one corresponding to
\emph{near-BPS} and the other to \emph{far from BPS} black hole
solutions. There were similarities and differences between the two
cases. In both cases taking the limit over the uplift of the $5d$
black hole solution to $10d$ IIB theory, we obtain a geometry
containing an $AdS_3\times S^3$ factor (or more generally a
$X_{M,J}\times S^3$ geometry, where $X_{M,J}$ is generically a
$3d$ stationary spacetime the Ricci curvature of which obeying
$R_{\mu\nu}=-\frac{1}{R^2}g_{\mu\nu}$). Therefore, there should be
a $2d$ CFT dual description. On the other hand, noting that the
starting $5d$ (or $10d$) geometry is a solution in the $AdS_5$ (or
$AdS_5\times S^5$) background there is a description in terms of
the dual $4d$ SYM theory. We identified the central charge of the
dual $2d$ CFT's in both cases and showed that the
Bekenstein-Hawking entropy of the original $5d$ solution, which is
the same as the Bekenstein-Hawking entropy of the $3d$ BTZ black
hole obtained after the limit, is correctly reproduced by the
Cardy formula of a $2d$ CFT, from which we identified the $L_0,
\bar L_0$ of the corresponding $2d$ CFT's in terms of the
parameters of the original $5d$ black hole. Matching of the
Bekenstein-Hawking entropy of the $5d$ and $3d$ black holes is a
strong indication that the near-horizon limit we are taking is
indeed a ``decoupling'' limit.

For the near-BPS case, the $2d$ description is essentially the
same as that of the D1-D5 system, modulo the
complication that our background corresponds not to a single ${\cal N} = ( 4,4)$ $2d$
CFT but a (continuous) collection of them, all of which have the
same $L_0,\bar L_0$ but different central charges. Nonetheless,
one can define an effective central charge for the system by
summing over the ``strip-wise'' $2d$ CFT descriptions.

For the far from BPS case, however, we have a different
situation; the conjectured $2d$ CFT description corresponds to a
set of D3 giants which have a deformed shape and as a result only
certain degrees of freedom on the giant theory survive our
(``$\alpha'\to 0$'') decoupling limit. In a sense, instead of
intersecting giants of the near-BPS case, at the extremal point
($\mu=\mu_c$) we are dealing with a (non-marginal) bound state of
giants. This could also be traced in the corresponding $6d$
gravity theory obtained from reduction of $10d$ IIB theory (see
Appendix \ref{Reduction-Appendix}). As discussed,  the two species
of the intersecting giants in the $6d$ language appear as strings
which are either electrically and/or magnetically charged under
the three-form $F_3$. The bound state of giants in the $6d$ theory
is then expected to appear as a usual ``$(Q_e, Q_m)$-string''. The
mass of this dyonic $(Q_e,Q_m)$-string state can be computed
working out the time-time component of the energy momentum tensor
of the system $T^0_0$ for the $AdS_3\times S^3$ configuration.
This has two parts, a cosmological constant piece and the part
which involves the three-form charges. The latter can be used to
identify
the mass squared of the $(Q_e,Q_m)$-string, which is%
\be\label{dyon-mass}%
M^2_{(Q_e,Q_m)}= T^{(6)}_s \left(N_e^2 \mathfrak{g}_s+
N_m^2 \mathfrak{g}_s^{-1}\right)
%=\frac{(q_2q_3)}{T^{(6)}_s}\sum_i\frac{\tilde q_i^2}{ q_i^2},%
\ee%
where $\mathfrak{g}_s=\langle X^{-2}\rangle $ is the ``effective''
$6d$ string coupling and $N_e,\ N_m$ are the number of electric
and magnetic strings and are related to $Q_e,\ Q_m$ as
\eqref{6d-charge}. \footnote{Note that with the action
\eqref{6d-action-section3.3} we are working in ``Einstein frame''
and the mass of fundamental string mass squared is
$T^{(6)}_s\mathfrak{g}_s$.}

To complete this picture one should in fact show that the $6d$
$(Q_e,Q_m)$-string discussed in section \ref{6d-analysis-section}
is indeed a BPS, stable configuration in the corresponding gravity
theory. Moreover, it is plausible to expect that our $6d$ gravity
description is a part of a new type of $6d$ gauged supergravity.
This $6d$ theory is expected to be a $U(1)^2$ ${\cal N}=(1,1)$
gauged SUGRA with the matter content (in the language of $6d$
${\cal N}=1$): one gravity multiplet, one tensor multiplet and two
$U(1)$ vector multiplets. This theory is nothing but a $6d$
version of the $d=4, \ d=5$ ``gauged STU'' models (\emph{e.g.} see
\cite{Cvetic:1999xp, Duff:1999rk}) and may be obtained from a
suitable generalization of the reduction we already discussed in
Appendix \ref{Reduction-Appendix}. The two $U(1)$ gauge fields
$A_i$ are coming from replacing $d\chi_i$ in reduction ans\"atz
\eqref{metric-reduc} with $d\chi_i+LA_i$. The details of this
reduction and construction and analysis of this ``$6d$ gauged
STU'' supergravity will be discussed in an upcoming publication
\cite{progress-6d}.

In section \ref{dual-gauge-section} we gave a description of both
the near-BPS and far from BPS cases in terms of specific sectors
of large $R$-charge, large engineering dimension operators. We
expect these sectors to be decoupled from the rest of the theory
since they also have a description in terms of a unitary $2d$ CFT.
The near-BPS case has features similar to the BMN sector. In this
case, however, the sector is identified with operators of $J_i\sim
N^{3/2}$, as opposed to  $J\sim N^{1/2}$ of BMN case. In the
far from BPS case the operators we are dealing with are far from
being BPS and their $R$-charge $J_i\ (i=2,3)$ scale as $N^2$.
Understanding these sectors in the $4d$ gauge theory and computing
their effective 't Hooft expansion parameters, namely effective 't
Hooft coupling and the planar-nonplanar expansion ratio, is an
interesting open question. From our analysis, however, we expect
there should be new ``double scaling limits'' similarly to the BMN
case. It is also desirable to give another supportive evidence for
the decoupling of these sectors by  counting degeneracy of the
states in both of these sectors in ${\cal N}=4$ SYM and matching
it with the Bekenstein-Hawking entropies computed here.

Here we focused on the two-charge $5d$ extremal black hole solutions
of $U(1)^3$ $5d$ gauged supergravity. The $U(1)^4$ $d=4$ gauged
supergravity has a similar set of black hole solutions
\cite{Cvetic:1999xp, Duff:1999rk, Leblond:2001gn}. Among them there
are three-charge extremal black holes. One can take the near-horizon
decoupling limits over these black holes to obtain $AdS_3\times S^2$
geometries. Again there are two possibilities, the near-BPS and
far from BPS cases, very much the same as what we found
here in the $5d$ case. Detailed analysis of these decoupling limits
is what we present in an upcoming work \cite{preparation-11d}.

\section*{Acknowledgments}
We would like to thank Mirjam Cvetic, Wafic Sabra, Ergin Sezgin,
Seif Randjbar-Daemi and Joan Simon for helpful discussions and
comments. CNG thanks the theory group at IPM for it's warm
hospitality during his visit there, when this project was initiated. The research of AEM is partly supported by TWASIC.

%\section*{Appendices}
\appendix
\section{Reduction $10\to 6$}\label{Reduction-Appendix}

Here we present some of the details of the computations for the
reduction of $10d$ IIB supergravity to the $6d$ gravity theory
discussed in section \ref{6d-analysis-section}. We start with the
following reduction ans\"atz for the $10d$ metric:
\begin{eqnarray}\label{metric-reduc}
ds^2_{(10)} &=& \mu_1\ ds^2_{(6)} + \frac{L^2}{\mu_1}\, \sum_{i=2,3}
X_i^{-1}(d\mu_i^2 + \mu_i^2 \, d\chi_i^2 )
 \end{eqnarray}
where $\chi_i$ range over $[0,2\pi]$, \be\label{mu1-def}
\mu_1^2=1-\mu_2^2-\mu_3^2\ , \ee and $ds^2_{(6)}$ is the $6d$
metric, $x$ denotes the $6d$ coordinates; \be X_2X_3=1,\ \qquad
X_2(x)\equiv X(x) \ee and $X(x)$ is the $6d$ scalar coming from the
reduction. As it is seen from \eqref{mu1-def}, $0\leq \mu_i\leq 1,\
i=2,3$. In what follows we will use Capital Latin indices
$M,N,P,\cdots$ for $10d$ coordinates, little Latin indices
$i,j,k,\cdots$ for the four-dimensional reduction manifold and Greek
indices $\mu,\nu,\cdots =0,1,\cdots, 5$ for the $6d$ spacetime.

In the absence of the scalar field, \emph{i.e.} when $X(x)=1$, the
reduction manifold ${\cal M}_4$ is simply a four-dimensional ball
of radius $L$ and hence is not ``compact'' in the topological
sense. Moreover, the $10d$ metric for this case has a curvature
singularity at $\mu_1=0$. We expect this singularity to be removed
once the stringy corrections are considered. The volume of the
${\cal M}_4$ is%
\be\label{V4}%
V_{{\cal M}_4}=(2\pi)^2 L^4 \int
\mu_2\mu_3 d\mu_2d\mu_3=\frac{\pi^2}{2} L^4\ . %
\ee%
Therefore the Newton constant of the $6d$ theory we are going to derive is%
\be\label{GN-6d-App}%
G^{(6)}_N=\frac{G^{(10)}_N}{V_{{\cal M}_4}}=\frac{G^{(10)}_N}{\frac{\pi^2}{2} L^4}\ . %
\ee

Besides the reduction ans\"atz for the metric we also need to give
the reduction ans\"atz for the other fields of the $10d$ IIB
theory. In our case we choose to turn off all the form fields and
the dilaton, except for the (self-dual) five-form of the IIB
theory which is reduced such that leads to a three-form $F_3$ in
$6d$. Explicitly,
\begin{eqnarray}\label{F5-reduc}
F_5 &=& X ( {\cal J}_+ \wedge F_{3+} + {\cal J}_- \wedge F_{3-} ), \\
\label{Jpm} {\cal J}_{\pm}&=& \frac{1}{X} d\mu_2^2 \wedge d\chi_2
\pm \,X\, d\mu_3^2 \wedge d\chi_3, \\ \label{F3} F_{3\pm} &=&
\frac{1 \pm \ast_6}{2} F_3.
\end{eqnarray}
where ${\cal J}_{\pm}$ (and $F_{3\pm}$) are the self-dual and
anti-self-dual two-form (three-form) fields on the ${\cal M}_4$
(in $6d$) and ${\cal J}_+\wedge {\cal J}_-$ is its volume form.
{}From the above reduction ans\"atz for five-form it is evidently
seen that $F_5$ is self-dual.

To show that the above reduction ans\"atz for metric and the
five-form really leads to a consistent reduction of IIB theory to
a six-dimensional theory we need to wok at the level of equations
of motion and show that set of $10d$ IIB equations of motion lead
to a consistent system of equations for a $6d$ gravity theory
coupled to a scalar $X$ and a three-form $F_3$. The IIB equations
of motion relevant to our case are (\emph{e.g.} see
\cite{Duff:1999rk})
\begin{eqnarray}
\textrm{e.o.m\ for metric:}&\quad R_{M\,N} = \frac{1}{96}\
(F^2_5)_{MN},\qquad (F^2_5)_{MN}\equiv
F_{M\,P_1\,P_2\,P_3\,P_4}\,F_N^{~~P_1\,P_2\,P_3\,P_4},
\\  \textrm{e.o.m\ for five-form:}&\qquad F_5 = \ast F_5, \quad d\,F_5 =
0.
\end{eqnarray}
It is notable that, as a result of self-duality of the five-form
\[F_5^2=F_{P_1\,P_2\,P_3\,P_4\,P_5}\,F^{P_1\,P_2\,P_3\,P_4\,P_5}=0,\]
and hence the equation of motion for metric also implies
$R=R_{MN}g^{MN}=0$, which in turn is the equation of motion for
constant dilaton for our ans\"atz.

Writing \eqref{F5-reduc} in components,%
\be\label{F5-reduc-comp} F_5 = \frac{1}{3 !} F_{3\,\mu\nu\rho} \,
d\mu_2^2\wedge d\chi_2 \wedge dx^\mu \wedge dx^\nu \wedge dx^\rho +
\frac{1}{3 !}X^2 (\ast F_3)_{\mu\nu\rho} \, d\mu_3^2\wedge
d\chi_3 \wedge dx^\mu \wedge dx^\nu \wedge dx^\rho, \ee%
it is evidently seen that the five-form equation of motion,
$dF_5=0$ implies the equations of motion for the three-form:%
\begin{eqnarray}\label{threeform-eom}%
d F_3 &=& 0\\ d\,(X^2 \, \ast F_3 )&=& 0.
\end{eqnarray}

The metric equations of motion, decomposes into three independent
set of equations; the $R_{\mu_2\mu_2},\ R_{\mu_3\mu_3}$ and the
$R_{\mu\nu}$ components. Computing the $10d$ Ricci tensor with the
ans\"atz \eqref{metric-reduc} we obtain
\begin{subequations}\label{Ricci-10d}\begin{align}
R_{\mu_2\mu_2}&=g^{(10)}_{\mu_2\mu_2}\frac{1}{\mu_1}\
\left(\frac{\Delta\,X}{2\,X} - \frac{\nabla X^2 }{2\,X^2}+
\frac{1}{L^2}\, (X -
X^{-1})\right)\\ %
\frac{R_{\mu_2\mu_2}}{g^{(10)}_{\mu_2\mu_2}}&=-\frac{R_{\mu_3\mu_3}}{g^{(10)}_{\mu_3\mu_3}}
\\%
R^{(10)}_{\mu\nu}&=R^{(6)}_{\mu\nu}+\frac{1}{L^2}\, (X +
X^{-1})\,g^{(6)}_{\mu\nu}-\frac{1}{X^2}\ \nabla_\mu X\nabla_\nu X.
\end{align}\end{subequations}
The right-hand-side of the metric equations of motion is also
computed as
\begin{subequations}\label{F5-MN}\begin{align}
(F^2_5)_{\mu_2\mu_2}&= g^{(10)}_{\mu_2\mu_2}\ \frac{16X^2}{\mu_1} F_{3\,\mu\nu\rho} F_3^{~\mu\nu\rho} \\
(F^2_5)_{\mu_3\mu_3}&= g^{(10)}_{\mu_3\mu_3}\ \frac{16X^2}{\mu_1}
(\ast F_3)_{\,\mu\nu\rho}(\ast F_3)^{~\mu\nu\rho}\\
(F^2_5)_{\mu\nu}&= 48X^2\left( F_{3\,\mu\rho\lambda}\,F_{3\,
\nu}^{~~\rho\lambda} +(\ast F_3)_{\mu\rho\lambda}\,( \ast F_3)_
\nu^{~~\rho\lambda} \right).\end{align}
\end{subequations}
Recalling that%
\be\label{F3-identity}%
\begin{split}
 F_{3\,\mu\rho\lambda}\,F_{3\, \nu}^{~~\rho\lambda} &=
(\ast F_3)_{\mu\rho\lambda}\,( \ast F_3)_
\nu^{~~\rho\lambda}+\frac13\,g_{\mu\nu}^{(6)}\,F_{3\,\alpha\rho\lambda}\,F_{3}^{\,\alpha\rho\lambda},\
\cr
F_{3\,\mu\rho\lambda}\,F_{3}^{\,\mu\rho\lambda} &= -(\ast F_3)_{\mu\rho\lambda}\,( \ast F_3)^{\mu\rho\lambda}\ , %
\end{split}%
\ee%
We see
that the $\mu_2\mu_2$ and $\mu_3\mu_3$ are consistent and become
identical. (The $\chi_i\chi_i$ components are hence identical
too.) This proves the consistency of our reduction ans\"atz.

In sum, the $10d$ equations of motion are all satisfied if $6d$
metric $g^{(6)}_{\mu\nu}$, $X$ and the three-form satisfy
\begin{subequations}\label{6d-eom}
\begin{align}
R^{(6)}_{\mu\nu}&=\frac{1}{X^2}\ \nabla_\mu X\nabla_\nu
X-\frac{1}{L^2}(X+X^{-1}) g^{(6)}_{\mu\nu}+X^2\left(
F_{3\,\mu\rho\lambda}\,F_{3\, \nu}^{~~\rho\lambda}-\frac16\,g_{\mu\nu}^{(6)}\,F_{3\,\alpha\rho\lambda}\,F_{3}^{\,\alpha\rho\lambda}\right)\\
&\frac{\Delta\,X}{2\,X} - \frac{\nabla X^2 }{2\,X^2}+
\frac{1}{L^2}\, (X - X^{-1})=\frac{X^2}{6}\
F_{3\,\mu\rho\lambda}\,F_{3}^{\,\mu\rho\lambda}\\
&d(X^2*F_3)=0, \qquad dF_3=0
\end{align}
\end{subequations}

The above equations of motion can be obtained from the $6d$
gravity action%
\begin{equation}\label{6d-action}
{\cal S}_6 = \frac{1}{16\pi G^{(6)}_N}\ \int d^6x\,\sqrt{-g^{(6)}}
\bigl[R^{(6)} - g^{\mu\nu}\frac{\nabla_\mu X \, \nabla_\nu X}{X^2}
+ \frac{4}{L^2}\, (X + X^{-1}) - \frac{X^2}{3}\, F_{3\,\mu\nu\rho}
F_3^{~\mu\nu\rho}\bigr].
\end{equation}
To bring the kinetic term into the canonical form one may define the
scalar field $\phi$ as
\[
X=e^{\phi}
\]
in which case the potential becomes $8\cosh\phi/L^2$.

\section{Conventions For Rotating BTZ and Conical
Spaces}\label{BTZ-Appendix}

In this appendix we give a brief review of the definitions of all
possible stationary $3d$ \emph{locally} $AdS_3$ spacetimes, obeying
$R_{\mu\nu}=-\frac{2}{R^2} g_{\mu\nu}$. The most generic solution is
of course the BTZ-type black hole%
\be\label{generic-metric-3d}%
\begin{split}%
 ds^2=R^2\biggl\{&-
 \frac{r^4+2(a^2_++a^2_-)r^2+(a_+^2-a_-^2)^2}{r^2} dt^2+ \frac{r^2}{r^4+2(a^2_++a^2_-)r^2+(a_+^2-a_-^2)^2} dr^2
 \cr &+
 r^2\left(d\phi+\frac{a_+^2-a_-^2}{r^2}dt\right)^2\biggr\}\ , %
\end{split} \ee%
 where $\phi\in [0,2\pi]$. Without loss of generality we
can always assume $a^2_+\leq a^2_-$. It
is useful to introduce to other parameters%
 \be\label{M-J,aa}
 a^2_+=-\frac{M+J}{4},\ \qquad a^2_-=-\frac{M-J}{4},\quad J\geq 0
 \ee%
We are then left with the following three possibilities
(\emph{e.g.} see \cite{David:2002wn})
\begin{itemize}
\item{ $a^2_+,\ a^2_- > 0$, corresponding to $M< -J$. In
this case we are generically dealing with a space with conic
singularity. The special case of $a_+=a_-=1/2$ corresponds to a
\emph{global} $AdS_3$. For the generic case $a_+=a_-=\gamma/2$,
where $J=0$, the conic space has the same line element as a global
$AdS_3$ but its $\phi$ coordinate is now ranging over
$[0,2\pi\gamma]$. In string theory for \emph{rational} values of
$\gamma$ and only when $\gamma<1$ the conical singularity could be
resolved \cite{Maldacena:2000dr}. For the general case when $a_+\neq
a_-$, the conical space can be resolved in string theory only when
$a_-^2$ is a rational number and  $0\leq a_-^2\leq 1/4$ (we are
assuming that $a_-^2\geq a_+^2$ and that $J\in \mathbb{Z}$). In
terms of $M, J$ that is%
\be\label{MJ-resolution-condition}%
-1\leq M-J\equiv -\gamma^2< -2J\ ,\qquad \gamma\in \mathbb{Q}, \
J\in \mathbb{Z}. \ee }
\item{ $a^2_+<0,\ a^2_->0$, corresponding to $-J<M<J$. The geometry is ill-defined and cannot
be made sense of in string theory.}
\item{ $a^2_+,\ a^2_-\leq 0$, corresponding to $M\geq J$, defines a rotating BTZ black hole of mass $M$
and angular momentum $J$ \cite{BTZ}. For this case the black hole
temperature (measured in units of $R$) is
\be\label{rotating-BTZ-Temp}\begin{split}%
T_{BTZ} =\frac{\sqrt{M^2-J^2}}{2\pi \rho_h},\qquad %
 \rho_h
=\frac{1}{2}\left(\sqrt{M+J}+\sqrt{M-J}\right).
\end{split}%
\ee%

The special case of $a_-=a_+$ (that is the $J=0$ case) corresponds
to static BTZ black hole. The $a_-=0$ (or equivalently $M=J$) case
corresponds to extremal rotating BTZ which has zero temperature and
finally the very special case of $a_-=a_+=0$ corresponds to massless
BTZ black hole.}
\end{itemize}
\begin{center}
\emph{To summarize the above, the cases with integer-valued $J$ and
when $M-J\geq -1$ are those which are sensible geometries in string
theory. For the $-1<M-J<0$ resolution of conical singularity in
string theory also demands $\sqrt{J-M}$ to be a rational number.}
\end{center}
As has been discussed in \cite{Maldacena:2000dr, Justin-Mandal,
Townsend, Henneaux} among the above cases $M\leq -J$ for any $M,
J$ and $M=J,\ M\geq 0$ can be supersymmetrized. For the $M \leq
-J$ case the solution becomes supersymmetric in a $3d$ gauged
supergravity which has at least two $U(1)$ gauge fields. In our
case to maintain supersymmetry one then needs to turn on the
Wilson lines of both of the $U(1)$ (flat-connection) gauge fields.
The two gauge fields which make the above metric supersymmetric
are then \cite{Maldacena:2000dr,Justin-Mandal, Townsend}
\be\label{SUSY-gauge-field}%
A^{(1)}=a_+(dt+d\phi),\qquad A^{(2)}=a_-(dt-d\phi)\ , %
\ee%
where $A^{(1)},\ A^{(2)}$ are the  flat connections of the two
$U(1)$'s. For $M= J,\ M\geq 0$ no gauge fields are needed to keep
supersymmetry. Among the supersymmetric configurations the
\emph{global} $AdS_3$, that is when $a_+=a_-=1/2$ keeps the
maximum supersymmetry the $3d$ theory has, with
\emph{anti-periodic} boundary conditions on fermions on the $\phi$
direction. The massless BTZ case, that is when $a_+=a_-=0$, as
well as the extremal BTZ (corresponding to $a_+^2=a_-^2>0 $) keep
half of the maximal supersymmetry but with \emph{periodic}
boundary conditions on the fermions on the $\phi$ direction
\cite{David:2002wn}. The conical spaces also keep half of maximal
supersymmetry.

\end{document}